\documentclass[prb,reprint,twocolumn,superscriptaddress,citeautoscript,amsmath,amssymb,floatfix]{revtex4-2}
\usepackage{orcidlink}
\usepackage{graphicx}
\usepackage{epstopdf}
\usepackage{subfigure}
\usepackage[percent]{overpic}
\usepackage{amsmath,amsthm,amssymb}
\usepackage{latexsym}
\usepackage{braket}
\usepackage[normalem]{ulem}
\usepackage{hyperref}
\usepackage[section]{placeins}
\usepackage[utf8]{inputenc}
\usepackage{soul}
\usepackage{dcolumn}
\usepackage{mathtools}
\usepackage{textcomp,gensymb} 
\usepackage{bm} 
\usepackage{float}
\usepackage{setspace}
\usepackage{array,booktabs}
\usepackage{changepage}
\usepackage{tikz}
\usepackage{contour}
\usepackage[left]{lineno}
\usepackage{blindtext}
\usepackage{setspace}
\usepackage[none]{hyphenat}
\usepackage{multirow}   
\usepackage{siunitx}
\definecolor{ashgray}{rgb}{0.7,0.75,0.71}
\definecolor{mspringgreen}{rgb}{0, 0.8, 0.1}
\definecolor{auburn}{rgb}{0.43, 0.21, 0.1}
\definecolor{ao(english)}{rgb}{0.0, 0.5, 0.0}
\definecolor{afw}{rgb}{0.95, 0.95, 0.96}
\definecolor{magnolia}{rgb}{0.97, 0.96, 1.0}
\definecolor{wsmk}{rgb}{0.96, 0.96, 0.96}

\newcommand{\ymb}{YbMnBi$_{2}$}

\newcommand{\cmb}{CaMnBi$_{2}$}

\newcommand{\cnmb}{Ca$_{\mathrm{1-}x}$Na$_x$MnBi$_{2}$}

\newcommand{\ba}{\mbox{\boldmath$a$}}
\newcommand{\bb}{\mbox{\boldmath$b$}}
\newcommand{\bc}{\mbox{\boldmath$c$}}

\newcolumntype{M}[1]{>{\centering\arraybackslash}m{#1}}

\newcolumntype{d}[1]{D{.}{.}{#1}}

\begin{document}
\bibliographystyle{apsrev4-2}
\title{Exploring spin-canting for Weyl states in topological semimetal CaMnBi$_2$}
\title{Does magnetism induce Weyl states in topological semimetal CaMnBi$_2$?}
\title{Electronically driven transition and magnetism in topological Weyl semimetal candidate CaMnBi$_2$}
\title{Magnetism and electronic structural transition in topological Weyl semimetal candidate CaMnBi$_2$}
\title{Magnetism and Peierls transition in Dirac semimetal CaMnBi$_2$}
\title{Magnetism and electronic structural transition in Dirac semimetal CaMnBi$_2$}
\title{Magnetism and electronic structural transition in Weyl semimetal candidate CaMnBi$_2$}
\title{Peierls distortion and magnetism in Dirac semimetal CaMnBi$_2$}
\title{Magnetism and Peierls distortion in Weyl semimetal candidate CaMnBi$_2$}
\title{Magnetism and Peierls distortion in Dirac semimetal CaMnBi$_2$}
\title{{Peierls Distortion and Magnetism in a Dirac Semimetal, CaMnBi$_2$}}
\title{{Peierls, Not Weyl: Bond-Order Distortion and Magnetism in the Dirac Semimetal CaMnBi$_2$}}
\title{{Peierls Bond-Order Distortion and Magnetism in a Dirac Semimetal, CaMnBi$_2$}}
\title{{Peierls Transition and Magnetism in a Dirac Semimetal: CaMnBi$_2$}}

\author{Aashish Sapkota\orcidlink{0000-0002-3572-1153}}
\affiliation {Condensed Matter Physics and Materials Science Division, Brookhaven National Laboratory, Upton, NY 11973, USA}
\affiliation{Ames National Laboratory, U.S. DOE, Iowa State University, Ames, Iowa 50011, USA}
\affiliation{Department of Physics and Astronomy, Iowa State University, Ames Iowa 50010, USA}
\author{Niraj Aryal\orcidlink{0000-0002-0968-6809}}
\affiliation{Condensed Matter Physics and Materials Science Division, Brookhaven National Laboratory, Upton, NY 11973, USA}
\author{Xiao Hu\orcidlink{0000-0002-0086-3127}}
\affiliation{Condensed Matter Physics and Materials Science Division, Brookhaven National Laboratory, Upton, NY 11973, USA}
\affiliation{Neutron Scattering Division, Oak Ridge National Laboratory, Oak Ridge, TN 37831, USA}
\author{Masaaki Matsuda\orcidlink{0000-0003-2209-9526}}
\affiliation{Neutron Scattering Division, Oak Ridge National Laboratory, Oak Ridge, TN 37831, USA}
\author{Yan Wu\orcidlink{0000-0003-1566-614X}} %
\affiliation{Neutron Scattering Division, Oak Ridge National Laboratory, Oak Ridge, TN 37831, USA}
\author{Guangyong Xu\orcidlink{0000-0003-1441-8275}}
\affiliation {NIST Center for Neutron Research, National Institute of Standards and Technology, Gaithersburg, MD 20899, USA}
\author{John M. Wilde\orcidlink{0000-0002-1753-3589}}
\affiliation{Ames National Laboratory, U.S. DOE, Iowa State University, Ames, Iowa 50011, USA}
\affiliation{Department of Physics and Astronomy, Iowa State University, Ames Iowa 50010, USA}
\author{Andreas Kreyssig\orcidlink{0009-0003-8474-2879}}
\affiliation{Ames National Laboratory, U.S. DOE, Iowa State University, Ames, Iowa 50011, USA}
\affiliation{Department of Physics and Astronomy, Iowa State University, Ames Iowa 50010, USA}
\affiliation{Institute for Experimental Physics IV, Ruhr-University Bochum, 44801 Bochum, Germany}
\author{Paul C. Canfield\orcidlink{0000-0002-7715-0643}}
\affiliation{Ames National Laboratory, U.S. DOE, Iowa State University, Ames, Iowa 50011, USA}
\affiliation{Department of Physics and Astronomy, Iowa State University, Ames Iowa 50010, USA}
\author{Cedomir Petrovic\orcidlink{0000-0001-6063-1881}}
\thanks{Present affiliation: Vin\v{c}a Institute of Nuclear Sciences, University of Belgrade, Mike Petrovi\'{c}a Alasa 12-14, 11351 Vin\v{c}a, Belgrade, Serbia; Shanghai Key Laboratory of Material Frontiers Research in Extreme Environments (MFree), Shanghai Advanced Research in Physical Sciences (SHARPS), Pudong, China.}
\affiliation{Condensed Matter Physics and Materials Science Division, Brookhaven National Laboratory, Upton, NY 11973, USA}
\author{John M. Tranquada\orcidlink{0000-0003-4984-8857}}
\author{Igor A. Zaliznyak\orcidlink{0000-0002-8548-7924}}
\email{zaliznyak@bnl.gov}
\affiliation{Condensed Matter Physics and Materials Science Division, Brookhaven National Laboratory, Upton, NY 11973, USA}

\date{\today}

\begin{abstract}
Dirac semimetals of the form $A$Mn$X_2$ ($A =$ alkaline-earth or divalent rare earth; $X =$ Bi, Sb) host conducting square-net Dirac-electron layers of $X$ atoms interleaved with antiferromagnetic Mn$X$ layers. {Hypervalent pnictogen square nets have long been predicted to host an electronically driven Peierls instability, whereas low-temperature anomalies in CaMnBi$_2$ near $T^{*}\sim 50$~K have previously been discussed in terms of possible magnetic canting, TRS breaking, and a resulting Weyl-semimetal (WSM) state. Here we use polarized and unpolarized neutron diffraction, x-ray diffraction, and density functional theory (DFT) to investigate this transition in single-crystal CaMnBi$_2$. We observe a coupled structural/magnetic symmetry-lowering transition at $T^{*}=46(2)$~K from a tetragonal lattice with C-type antiferromagnetism to an orthorhombic phase with unit-cell doubling along the $c$ axis, minimal impact on magnetism, and no measurable uniform spin canting ($\lesssim 2^\circ$). Superlattice peak intensities and the lattice distortion reveal a continuous second-order transition governed by a single primary order parameter. The refined displacement pattern corresponds to a zigzag bond-order-wave (BOW) modulation of in-plane Bi--Bi bonds, providing direct diffraction evidence for a Peierls-type instability of the Dirac-electron Bi layer anticipated by Hoffmann and co-workers [W.~Tremel and R.~Hoffmann, \textit{J. Am. Chem. Soc.} \textbf{109}, 124 (1987); G.~A.~Papoian and R.~Hoffmann, \textit{Angew. Chem. Int. Ed.} \textbf{39}, 2408 (2000)]. These results identify the $T^{*}$ anomaly as an intrinsic Dirac-electron-driven Peierls/BOW transition of the Bi square net.
}

\end{abstract}

\maketitle

\section{Introduction}
\label{Int}

Extensive research on topological quantum materials (TQMs) over the past decade has uncovered a range of novel electronic phenomena, providing platforms for controlling symmetry, topology, and electronic order, with potential technological applications in quantum computation, infrared sensors, chemical catalysis, and spintronics \cite{Khodas_2009,Wehling_2014,Katmis_Nature2016,Masuda_2016,Leslie_2018}. The initial impetus came from the discovery of Dirac fermions in graphene \cite{Novoselov_2005,Leslie_2018}, which spurred searches for Dirac-like quasiparticles and topological band structures in layered quasi-two-dimensional (2D) and bulk three-dimensional (3D) materials. Beyond topology itself, an increasingly important theme is how these Dirac-derived electronic states couple to collective instabilities, including magnetism, density-wave, bond-order, and nematic distortions, that lower crystal symmetry and reconstruct the low-energy bands.

One particularly interesting class of TQMs is topological semimetals (TSMs), which are defined by symmetry-protected band touching points (nodes) in the vicinity of the Fermi energy and can be considered a 3D analogue of the prototypical TQM graphene \cite{Wang_2011,Wehling_2014,Burkov_2016,Yan_2017,Klemenz_2019}. Depending on the character of the topological band crossing, TSMs are further classified into different families.
The point-like crossing of spin-degenerate bands gives rise to four-fold degenerate Dirac nodes in Dirac semimetals (DSMs), while lifting the spin degeneracy can yield two-fold crossing points of spin-split bands in Weyl semimetals (WSMs). Because the low-energy electronic structure is controlled by symmetry-protected crossings with a low density of states (DOS), even subtle symmetry breaking can qualitatively reconstruct Dirac states, making TSMs a natural platform for electronically driven instabilities.

{One such remarkable electronic instability is the Peierls transition, originally predicted by Peierls for one-dimensional metals at half-filling, leading to an intrinsic breakdown of the metallic state \cite{Peierls_book_1955}. Later, similar Peierls instabilities were predicted for Dirac electrons in pnictogen square nets near half-filling \cite{TremelHoffman_JACS1987,PapoianHoffman_AngewChem2000}. This instability is remarkable because at its core it is entirely kinematic and thus profoundly generic: it relies only on the Pauli exclusion principle and electron-lattice coupling and requires no interaction between the mobile electrons. At half-filling, a distortion doubling the unit cell opens a gap at the Fermi level via bond modulation and lowers the energy of the filled band, yielding an insulating state.  An analogous mechanism underlies spin-Peierls dimerization in spin-1/2 chains, describable in terms of neutral Jordan-Wigner fermions \cite{CrossFisher_PRB1979}.  This differs markedly from a charge-density-wave instability driven primarily by electron-electron Coulomb interactions. For two-dimensional Dirac electrons, several bond-order patterns reconstructing the Fermi surface were predicted \cite{TremelHoffman_JACS1987,PapoianHoffman_AngewChem2000}. }

{On the other hand, a} Weyl semimetallic state (WSS), hosting Weyl fermions, the chiral anomaly, and Fermi arcs, can arise when a Dirac semimetal is perturbed by breaking either spatial inversion symmetry (IS) or time-reversal symmetry (TRS), yielding spin-split bands \cite{Leslie_2018,Wang_2023}. The latter route is often sought in magnetic topological semimetals. At the same time, symmetry lowering driven by electronic and electron-lattice interactions, such as density-wave or Peierls bond-order instabilities, provides an alternative and broadly relevant mechanism for reconstructing Dirac bands without TRS breaking or a net ferromagnetic moment, and can, in principle, compete or intertwine with WSS physics.

\begin{figure*}[t!]
\centering
\includegraphics[width=2.\columnwidth]{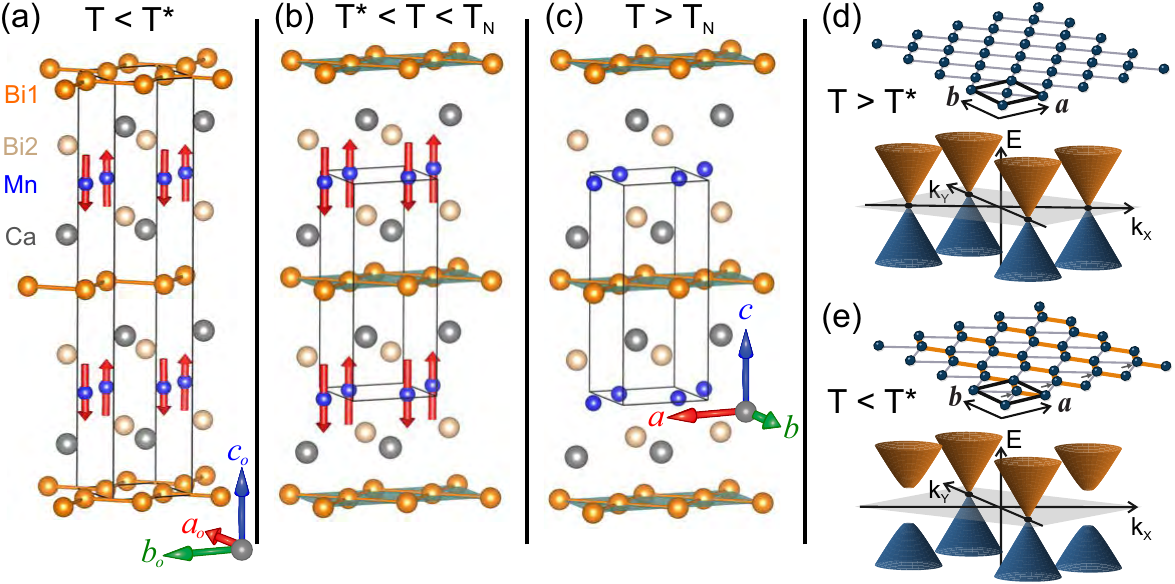}
\caption{Crystal, magnetic, and electronic structures of \cmb\ across three relevant temperature regimes.
(a) Orthorhombic C-type antiferromagnetic $Pc^\prime m^\prime n^\prime$ phase with nuclear $Pcmn$ symmetry at $T<T^{*}$.
(b) Tetragonal C-type antiferromagnetic $P4^\prime/n^\prime m^\prime m$ phase with space group $P4/nmm$ for $T_\mathrm{N}>T>T^{*}$.
(c) Tetragonal paramagnetic $P4/nmm$ phase for $T>T_\mathrm{N}$.
Although $Pnma$ is the conventional setting for orthorhombic 112 compounds, we use the non-standard $Pcmn$ setting to preserve the axis orientation and labeling of the high-temperature $P4/nmm$ tetragonal structure.
The shaded regions in panels (b) and (c) indicate the Bi square lattice that hosts the Dirac bands, which develops a zigzag bond-order-wave distortion in the orthorhombic $Pcmn$ phase shown in panel (a).
(d) Schematic undistorted Bi square net (top) and corresponding Dirac band structure at half filling (bottom); the shaded plane denotes the Fermi energy.
(e) Schematic Peierls/BOW zigzag distortion of the Bi square net (top) and the associated gap opening at two of the four Dirac cones (bottom).}
\label{CMS}
\end{figure*}

The 112 ternary pnictides with general formula $A$Mn$X_2$ ($A =$ alkaline-earth or divalent rare-earth metal; $X =$ Bi, Sb) provide an exemplary magnetic TSM platform \cite{Wang_2011,Wang_2016,Guo_PRB2014,Sapkota_2020,Soh_PRB2021,Hu_PRB2023}. Their crystal structure comprises square-net layers of $X$ atoms hosting topological Dirac fermions, interleaved with Mn$X$ planes and $A$ cation layers [Fig.~\ref{CMS}]. The Mn atoms form a square lattice with antiferromagnetic (AFM) order near room temperature [Fig.~\ref{CMS}(b)]. 
{In this family, weak ferromagnetism arising from uniform canting of Mn moments can generate a TRS-breaking in-plane ferromagnetic (FM) moment and has been discussed as a route to a type-II Weyl state. However, the same square-net chemistry also renders these materials susceptible to Peierls-type reconstruction of the Dirac bands associated with an electronically driven bond disproportionation in the $X$ layers that lowers the crystal symmetry and reconstructs the low-energy electronic structure \cite{TremelHoffman_JACS1987,PapoianHoffman_AngewChem2000}.} 

In CaMnBi$_2$, which we study here, an unexpected transition was revealed by measurements of the in-plane resistivity, which shows a cusp with a slight upturn upon cooling near 50~K \cite{Wang_2012,He_2012}. An optical study reported a decrease in the Drude weight below this transition \cite{Yang_2020}, while magnetic torque measurements, combined with \emph{ab initio} DFT calculations and analysis of the optical response were interpreted as evidence for a $\sim 10^\circ$ in-plane canting of the Mn moments \cite{Yang_2020}.
{Testing this hypothesis and resolving the origin of the $T^{*}$ anomaly were the original motivations for the present study; as we show below, the data instead point to an intrinsic Dirac-electron-driven symmetry-lowering instability of the Bi square-net layer.}

{Central to our results and to the electronic behavior of the 112 family more broadly is the Peierls-type instability of hypervalent pnictogen square nets analyzed by Hoffmann and collaborators \cite{TremelHoffman_JACS1987,PapoianHoffman_AngewChem2000}. For a square net of $X$ (Sb or Bi) atoms, the electronic states near the Fermi level are dominated by bands derived from the nearly half-filled $p_x$ and $p_y$ orbitals; their crossings form four symmetry-related Dirac cones [Fig.~\ref{CMS}(d)].
When the Fermi energy lies at or near these Dirac nodes, close to half filling, a Peierls/BOW distortion can lower the energy of the occupied electronic states by gapping the Dirac points. In \cmb, this distortion shifts the centering Bi atoms, producing a zigzag modulation of the in-plane Bi--Bi bonds and lowering the lattice symmetry from tetragonal to orthorhombic; the equilibrium displacement reflects the balance between electronic energy gain and the cost of lattice elastic strain. This bond modulation enables hybridization of the $p_x$ and $p_y$ bands and reconstructs the Dirac bands by gapping two of the four Dirac cones [Fig.~\ref{CMS}(e)], providing a natural microscopic origin for the partial Fermi-surface gapping associated with the anomaly at $T^{*}$.} 
%

While \textcite{TremelHoffman_JACS1987,PapoianHoffman_AngewChem2000} cited SrZnSb$_2$ as the only example of Peierls-distorted orthorhombic structure with zig-zag Sb chains, recent studies of $A$MnSb$_2$ compounds have identified similar distortions in many other cases: $A=$ Sr \cite{You_CurrApplPhys2019,Zhang_PRB2019,Ning_PRB2024}, Ca \cite{Brechtel_JLessCommMet1981,He_PRB2017,Qiu_PRB2018,Rong_PRB2021}, Ba \cite{Sakai_PRB2020,Liu_NatComm2021,Yoshizawa_PRB2022}, Eu \cite{Wilde_PRB2022}, and Yb \cite{Bhoi_PRB2025}, correcting earlier reports of the undistorted tetragonal structure with ideal square nets for $A=$ Ba \cite{Liu_SciRep2016,Ramankutty_SciPost2018} and Yb \cite{Soh_PRB2021,Tobin_PRB2023}. Moreover, the zigzag distortion was found to persist even in a high-entropy mixed-anion compound, Ba$_{0.38}$Sr$_{0.14}$Ca$_{0.16}$Eu$_{0.16}$Yb$_{0.16}$MnSb$_2$ \cite{Laha_NatComm2024}, showing that the Peierls-distorted orthorhombic structure is robust and common among Sb compounds of the 112 family.

On the other hand, no structural distortion has been observed in YbMnBi$_2$ \cite{Wang_2016, Soh_2019, Sapkota_2020}, EuMnBi$_2$ \cite{Masuda_2016, Borisenko_2019}, SrMnBi$_2$, or CaMnBi$_2$ \cite{Guo_PRB2014, He_PRB2017}, all of which were previously reported to retain the ideal square-net tetragonal ($P4/nmm$) structure.
{These results indicate that whereas the zigzag BOW motif associated with the square-net Peierls tendency is robust and widespread in Sb-based 112 compounds, the corresponding instability is weaker in Bi-based members and may emerge only as a low-temperature, electronically driven transition.}

This absence of distortion in Bi compounds is, in principle, consistent with a reduced Peierls instability in the heavier pnictogen (Bi) layers. As discussed in Ref.~\onlinecite{PapoianHoffman_AngewChem2000}, this trend can be rationalized by weaker $sp$ hybridization in higher-$Z$ elements, which suppresses the electronic instability; the same analysis also predicted that a less electronegative $A$-site cation further disfavors distortion.
Nevertheless, a slight structural distortion was recently reported in BaMnBi$_2$, with an orthorhombicity about 10\% of that in BaMnSb$_2$ \cite{Kondo_2021}, revising earlier reports of an undistorted square-net configuration \cite{Ryu_SciRep2018}.

In this paper, we present polarized and unpolarized neutron diffraction along with x-ray diffraction measurements on single crystals of \cmb. These complementary probes reveal a coupled structural and magnetic phase transition at $T_\mathrm{s,m} = 46(1)$~K, corresponding to the previously reported resistivity anomaly.
{Below $T_{\mathrm{s,m}}$, \cmb\ adopts an orthorhombic $Pcmn$ (standard setting: $Pnma$) structure with unit-cell doubling along $c$, and the refined displacement pattern forms a zigzag BOW in the Bi square net, the diffraction signature expected for the Peierls mechanism proposed by Hoffmann and collaborators \cite{TremelHoffman_JACS1987,PapoianHoffman_AngewChem2000}.} 

The magnetic order remains essentially C-type AFM. Although the orthorhombic magnetic symmetry allows in-plane canting within each MnBi layer, the $c$-axis doubling requires any in-plane ferromagnetic component to alternate sign between adjacent layers, preserving bulk TRS in the semimetallic Bi layers. Consistent with this, we observe no discernible \emph{uniform} canting; if present, it lies below our detection limit of $\lesssim 2^\circ$.

Our central result is the experimental identification of an intrinsic Dirac-electron-driven Peierls/BOW instability in a Bi-based square-net Dirac semimetal, and its subtle influence on the antiferromagnetism in the intervening MnBi layers.
{Our results establish a direct experimental link between square-net Dirac semimetals and Peierls/BOW physics, providing an electronically driven, compositionally tunable route to symmetry breaking and low-energy electronic band reconstruction in a widely studied topological materials family.}

The remainder of this paper is organized as follows. Background information on the structure and physical properties of \cmb\ is provided in Sec.~\ref{BICMB}. Experimental procedures are described in Sec.~\ref{ED}, followed by detailed analysis of the neutron and x-ray diffraction results in Sec.~\ref{RnA}. Sec.~\ref{DFT} presents the DFT calculations. {The implications of these results are discussed in Sec.~\ref{Sec:Discussion}, followed by a brief summary and conclusions in Sec.~\ref{Sec:Summary}.} Additional supporting experimental data and analysis details are provided in the Appendices.

\section{Structure and Physical Properties of C\lowercase{a}M\lowercase{n}B\lowercase{i}$_2$}
\label{BICMB}

\cmb\ is a Dirac semimetal characterized by strongly anisotropic Dirac dispersion, with Dirac points located near the Fermi energy \cite{Wang_2012, Lee_2013}. Its crystal structure, with tetragonal $P4/nmm$ symmetry at room temperature, was first reported in 1980 based on measurements on a polycrystalline sample \cite{Brechtel_1980}. All recent measurements on both single-crystal and powder samples at room temperature confirm the structure to be $P4/nmm$ \cite{Wang_2012, He_2012, Guo_PRB2014, Wang_2016, Zhang_InorgChem2022}. For this crystal structure, Bragg reflections of the $(H, K, 0)$ type are allowed for $H + K = 2n$, and of the $(H, 0, 0)$ type for $H = 2n$, where $n$ is an integer.

In-plane electronic transport measurements \cite{Wang_2012, He_2012, Guo_PRB2014} reveal metallic behavior, but with a relatively high in-plane residual resistivity of $\gtrsim 20~\mu \Omega \cdot\mathrm{cm}$ at $T \rightarrow 0$~K, indicating that \cmb\ is a moderately bad metal \cite{He_2012}. The temperature dependence of magnetic susceptibility, $\chi(T)$, exhibits a feature associated with AFM ordering, with a reported N\'{e}el temperature $T_\mathrm{N}$ ranging from 250~K to 300~K across different studies \cite{Wang_2012, He_2012, Guo_PRB2014}. No clear signatures of the AFM transition are observed in resistivity measurements. However, at a lower temperature around $T^{*} \sim 50$~K, Refs.~\citenum{Wang_2012} and \citenum{He_2012} reported a hump- or cusp-like anomaly in resistivity, interpreted as a manifestation of Fermi surface instability and reconstruction of the Fermi surface resulting from uniform spin canting \cite{He_2012, Corasaniti_2019, Yang_2020}. A subtle, plateau-like feature in $\chi(T)$ is also observed near $T^{*}$ \cite{He_2012}. It was also found that Na substitution at the Ca site enhances the resistive anomaly, with $T^{*}$ in \cnmb\ increasing to $\approx 75$~K and $\approx 100$~K for $5\%$ and $9\%$ Na substitution, respectively \cite{Corasaniti_2019,Yang_2020}.

Previous neutron powder diffraction measurements on a polycrystalline sample of \cmb\ identified a N\'{e}el-type (also known as C-type) AFM ordering described by the $P4^\prime /n^\prime m^\prime m$ magnetic space group below $T_\mathrm{N} \approx 300$~K, with an ordered Mn magnetic moment of $\mu_{\mathrm{Mn}} \approx 3.7 \mu_B$ at $T = 10$~K \cite{Guo_PRB2014}. In this magnetic structure, Mn moments are aligned along the easy $c$-axis with the nearest-neighbor moments in the $ab$-plane aligned antiferromagnetically and those along the $c$-axis aligned ferromagnetically. The same magnetic structure and ordered moment were later confirmed in a single-crystal diffraction study, although a slightly lower N\'eel temperature of $T_\mathrm{N} \approx 264$~K was reported \cite{Rahn_PRB2017}. The magnetic propagation vector is $\bm{\tau} = (1, 0, 0)$; thus, for a crystal aligned in the $(H, 0, L)$ horizontal scattering plane, magnetic Bragg peaks are expected at reciprocal lattice vectors of the form $(1 + 2n, 0, m)$, where $n$ and $m$ are integers \cite{Zaliznyak_2017, Rahn_PRB2017}.

If the antiferromagnetic structure were uniformly canted, producing a ferromagnetic component in the $ab$ plane, magnetic intensity should appear at these and other nuclear Bragg positions. Although this possibility was not specifically investigated, previous neutron-diffraction studies reported no evidence of the putative spin canting inferred from bulk measurements in the low-temperature magnetic structure.

The implications of spin canting and the resulting in-plane ferromagnetism are intriguing, as they raise the possibility of a Weyl semimetallic state in \cmb\ or its doped variants. However, the magnetic torque measurements from which a surprisingly large $\sim 10^\circ$ spin canting was inferred in Ref.~\citenum{Yang_2020} were performed on a bulk sample and are therefore susceptible to contributions from ferromagnetic inclusions, impurity phases, or crystal imperfections. Indeed, the authors observed ferromagnetic moments of comparable magnitude both in the $ab$-plane and along the $c$-axis, attributing the latter to the presence of vacancies \cite{Yang_2020, Liu_NatMat2017}. Moreover, torque measurements are sensitive to magnetic field misalignment, which can introduce significant bias given the subtle effect of the putative in-plane FM moment on the measured angular dependence of magnetic torque \cite{Yang_2020}. Finally, the measurements were conducted in applied magnetic fields of a few Tesla, which may enhance misalignment effects and could also modify the zero-field magnetic structure, potentially leading to incorrect conclusions about canted magnetism.

Neutron diffraction measurements are essential for determining whether spin canting and the resulting weak ferromagnetism appear in CaMnBi$_2$ below the resistive-anomaly temperature $T^{*}$. A canting angle of $\sim 10^{\circ}$, as reported in Ref.~\citenum{Yang_2020}, would be readily detectable by neutron scattering \cite{Zaliznyak_2017,Soh_2019}. Here, we use single-crystal neutron diffraction to track the evolution of the magnetic and lattice structures of CaMnBi$_2$ across $T^{*}$. While our data rule out the large canting and weak ferromagnetism inferred from bulk torque measurements, they reveal an intriguing structural transition: the Bi square net undergoes a distortion that breaks it into zigzag chains, consistent with the two-dimensional Peierls-type {bond-/charge-density-wave (BOW/CDW) predicted by Hoffmann and co-workers \cite{TremelHoffman_JACS1987,PapoianHoffman_AngewChem2000}, as illustrated in Fig.~\ref{CMS}.}

\section{Experimental Details}
\label{ED}

Single crystals of \cmb\ were grown using a high-temperature self-flux method, following the procedure described in Ref.~\citenum{Wang_2012}.
Stoichiometric mixtures of Ca (99.99\%), Mn (99.9\%), and excess Bi (99.99\%) with a ratio of Ca:Mn:Bi = 1:1:9 were sealed in a quartz tube, heated to 1050$^\circ$C, and then cooled to 450$^\circ$C, where the crystals were decanted. The resultant crystals are plate-like and the basal plane of a cleaved crystal is the crystallographic $ab$ plane.
Electrical transport measurements were performed using a conventional four-wire method in a Quantum Design Physical Property Measurement System (PPMS-9). Magnetization measurements were carried out in a Quantum Design Magnetic Property Measurement System (MPMS).

For neutron diffraction measurements, a 160~mg plate-like crystal was mounted on an aluminum plate and aligned in the $(H, 0, L)$ horizontal scattering plane. The sample holder was attached to the cold finger of a closed-cycle cryostat, enabling temperature control from 5 to 315~K. Unpolarized neutron scattering measurements were performed on the SPINS triple-axis spectrometer at the NIST Center for Neutron Research (NCNR) with a fixed final energy of $E_\mathrm{f} = 5$~meV and horizontal beam collimations of G–80$^\prime$–S–80$^\prime$–open (G = guide, S = sample). Two beryllium filters, placed before and after the sample, were used to suppress higher-order contamination from the monochromator.

Polarized neutron diffraction measurements were carried out on the HB-1 thermal neutron triple-axis spectrometer (TAS) at the High Flux Isotope Reactor (HFIR), Oak Ridge National Laboratory, using the same crystal previously measured with unpolarized neutrons. A Heusler monochromator and analyzer were employed for polarization analysis, and measurements were conducted in the \textbf{Q}$\parallel$\textbf{P} configuration, where \textbf{Q} is the wave vector transfer and \textbf{P} is the neutron polarization. The horizontal beam collimations were 48$^{\prime}$-80$^{\prime}$-S-60$^{\prime}$-240$^{\prime}$, with a fixed incident neutron energy of 13.5~meV. Contamination from higher-order reflections was effectively suppressed using pyrolytic graphite (PG) filters. A flipping ratio of 16 was measured at nuclear Bragg reflections.

For improved structural refinement, additional unpolarized neutron diffraction measurements were performed in four-circle mode on the DEMAND (HB-3A) diffractometer at HFIR. The same crystal previously used in the triple-axis measurements was aligned in the $(H, K, 0)$ horizontal scattering plane, and neutrons with a wavelength of 1.003~\AA, selected by a Si (331) monochromator, were used. Magnetic and nuclear structures were refined using the FullProf Suite.

Single-crystal x-ray diffraction measurements were performed on a four-circle diffractometer using Cu $K_{\alpha1}$ radiation from a rotating anode x-ray source at Ames National Laboratory, using a germanium (111) monochromator. Measurements were conducted on a 100~mg single crystal mounted on a flat copper sample holder, which was attached to the cold finger of a closed-cycle helium refrigerator to enable temperature-dependent studies. Three beryllium domes were used as the vacuum shroud, heat shield, and sample enclosure. The innermost dome, which housed the sample, was filled with a small amount of helium exchange gas to improve thermal contact.

Except for the low-temperature data used in the refinement of the $Pcmn$ ($Pnma$) crystal structure, most of the single-crystal data---including that shown in the figures---are indexed using the high-temperature tetragonal $P4/nmm$ lattice. The momentum transfer, $\mathbf{Q} = (H, K, L) = (2\pi/a) H \hat{\mathbf{i}} + (2\pi/b) K \hat{\mathbf{j}} + (2\pi/c) L \hat{\mathbf{k}}$, is expressed in this setting, with unit cell parameters $a = 4.50$~\AA\ and $c = 11.07$~\AA\ \cite{Guo_PRB2014, Rahn_PRB2017}.

\section{Experimental Results and Analysis}
\label{RnA}

\subsection{Bulk Properties}
\label{BulkProp}
%
\begin{figure}[]
\centering
\includegraphics[width=0.9\columnwidth,clip]{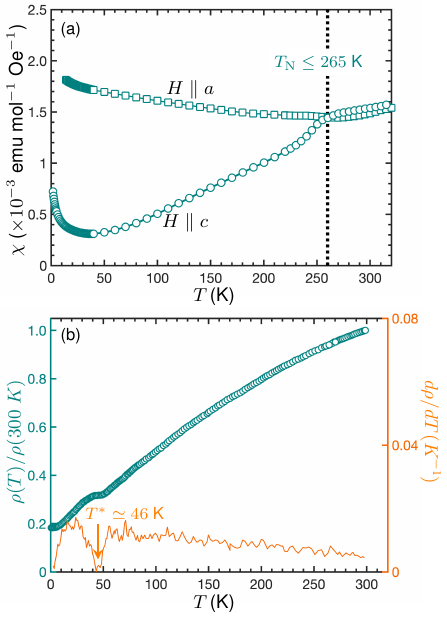}
\caption{Temperature dependence of the magnetic susceptibility, $\chi(T) = M(T)/H$, (a) and normalized resistivity, $\tilde{\rho} = \rho(T)/\rho(\mathrm{300~K})$ (left scale), and its derivative, $d\tilde{\rho}/dT$ (right scale), (b) of \cmb. Panel (a) shows zero-field-cooled susceptibility measured with an applied field of $\mu_0 H = 1$T for $\mathbf{H} \parallel \mathbf{c}$ and $\mathbf{H} \parallel \mathbf{a}$
Characteristic temperatures $T_\mathrm{N}$, associated with magnetic ordering, and $T^*$, discussed as a possible spin reorientation in Ref.~\citenum{Corasaniti_2019}, are indicated in the susceptibility and resistivity plots, respectively.}
\label{MR}
\end{figure}
%

To verify that the physical properties of the \cmb\ single crystals synthesized for this study are consistent with previously reported results, bulk characterization via magnetization and transport measurements was performed on selected crystals from the same batch used in our neutron and x-ray diffraction experiments. Figures~\ref{MR}(a) and \ref{MR}(b) show the temperature dependence of the magnetic susceptibility, $\chi(T) = M(T)/H$, and resistivity, $\rho(T)$, respectively.

Consistent with previous studies \cite{Wang_2012, He_2012}, the temperature dependence of the $c$-axis susceptibility, $\chi_{c}(T)$, exhibits a kink near 265~K, marking the onset of antiferromagnetic ordering [Fig.~\ref{MR}(a)]. Furthermore, the divergence between $\chi_{c}$, which decreases below $T_{\mathrm{N}}$, and $\chi_{ab}$, which remains nearly constant, confirms that the ordered Mn moments align along the easy $c$-axis, as previously reported (the small Curie-like upturn at low temperatures is likely due to a minor contribution from paramagnetic impurities in the sample).

The $T_{\mathrm{N}}$ of our crystals is approximately 10\% lower than $T_{\mathrm{N}} = 300(5)$~K reported by \textcite{Guo_PRB2014}, indicating notable sensitivity to synthesis conditions. Refinement of site occupancies from neutron diffraction (Appendix~\ref{App:HB3A_refinement}) shows that our crystal is stoichiometric within $\approx 1.4\%$,  comparable to the electron-probe microanalysis results of \textcite{Guo_PRB2014} for their samples, indicating that stoichiometric variations between samples are indeed small. Nevertheless, even minor deviations in stoichiometry can noticeably shift the Fermi level of the Dirac-electron Bi layers, which mediate magnetic coupling between the antiferromagnetic Mn–Bi layers \cite{Sapkota_2020, Hu_PRB2023}, thereby significantly affecting $T_{\mathrm{N}}$.

The resistivity measured on a \cmb\ single crystal with in-plane current [Fig.\ref{MR}(b)] shows no clear signature of the AFM transition near 265~K, consistent with Ref.~\onlinecite{Wang_2012} where no anomaly was detected for current along the $c$-axis either. In contrast, a clear cusp- or hump-like anomaly is evident at $T^{*} \approx 46$~K, in agreement with earlier reports \cite{Wang_2012, He_2012, Yang_2020}. This anomaly was absent in samples of \textcite{Guo_PRB2014} with higher $T_{\mathrm{N}}$, but becomes more pronounced and shifts to higher temperature in Ca$_{1-x}$Na$_{x}$MnBi$_2$ ($x = 0.03$ and $0.05$) \cite{Corasaniti_2019}, where Na doping lowers the Fermi level. It is therefore plausible that higher filling of the Bi Dirac-electron bands is responsible for both the elevated $T_{\mathrm{N}}$ and the absence of the resistivity anomaly in the samples of Ref.~\onlinecite{Guo_PRB2014}.

\subsection{N\'eel-type Antiferromagnetic Order}
\label{NMO}
\begin{figure}[t!]
\centering
\includegraphics[width=1.\columnwidth]{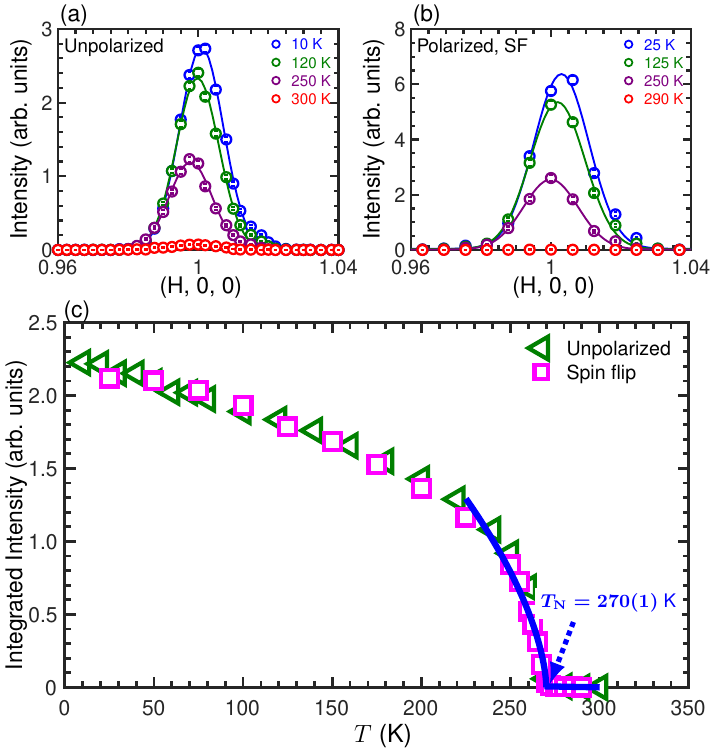}
\caption{Evolution of the (1, 0, 0) magnetic Bragg peak with temperature from unpolarized (a) and polarized (spin-flip channel) (b) neutron diffraction measurements. The slight red-shift in peak position upon warming results from thermal expansion. (c) Temperature dependence of the integrated magnetic intensity from unpolarized (triangles) and polarized (squares) measurements. The magnetic intensity for the unpolarized data was obtained by subtracting the average intensity above $T_\mathrm{N}$, treated as a non-magnetic background. The two datasets are cross-normalized to overlay. The solid blue line is a fit to a critical-type power law, $I \sim (T_{\mathrm{N}} - T)^{2\beta}$.}
\label{Fig:Neel_OP}
\end{figure}
%
As noted above, magnetic Bragg peaks corresponding to C-type AFM order in \cmb\ are expected at $(1 + 2n, 0, m)$ reciprocal lattice positions, where $n$ and $m$ are integers. Figures~\ref{Fig:Neel_OP}(a) and (b) show that the (1, 0, 0) peak, where nuclear scattering is forbidden, emerges below $T_\mathrm{N}$ in both the unpolarized and spin-flip (SF) channels of the polarized neutron measurements. Enhanced intensity was also observed at other $(1, 0, L)$ reflections, which contain both nuclear and magnetic scattering contributions. No magnetic signal was detected below $T_\mathrm{N}$ at $(0, 0, L)$-type reflections, consistent with Mn moments aligned along the $c$-axis (see also the discussion in Sec.~\ref{UPNSC}). The temperature dependence of the integrated intensity of the (1, 0, 0) peak from both polarized and unpolarized measurements is shown in Fig.~\ref{Fig:Neel_OP}(c). A power-law fit to the data near the transition yields $T_\mathrm{N} = 270(1)$~K and a critical exponent $\beta = 0.31(3)$, close to the value $\beta \approx 0.327$ for the three-dimensional (3D) Ising model. 

\subsection{Absence of Measurable Uniform Spin Canting}
\label{UPNSC}
\begin{figure}[]
\centering
\includegraphics[width=1.\columnwidth]{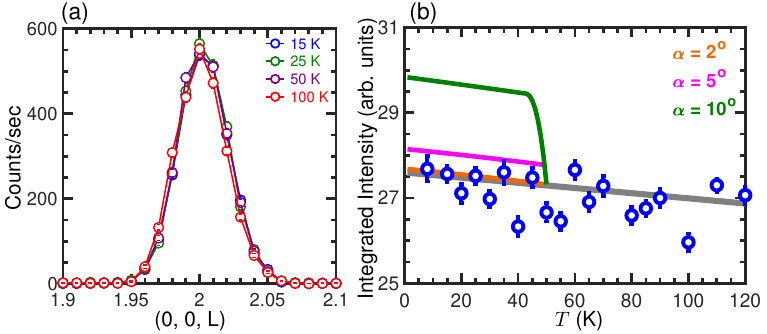}
\caption{(a) The $(0, 0, 2)$ lattice Bragg peak measured at four different temperatures above and below $T^{*}$, showing no apparent enhancement at low temperature. (b) Temperature dependence of the integrated intensity of the $(0, 0, 2)$ peak up to 120~K. The gray line represents a linear fit to the measured data, while the colored lines indicate the expected changes due to a ferromagnetic contribution, calculated for canting angles $\alpha = 2^\circ$ (orange), $5^\circ$ (magenta), and $10^\circ$ (green). }
\label{00L_FM}
\end{figure}
%

We next investigated the magnetic and lattice responses associated with the resistive anomaly at $T^{*} \approx 50$~K [Fig.~\ref{MR}(b)], which has previously been attributed to uniform canting of Mn moments away from the $c$-axis \cite{Wang_2011, He_2012, Corasaniti_2019, Yang_2020}, using unpolarized neutron diffraction. Such canting introduces a weak FM component in the $ab$-plane, which contributes magnetic intensity to the $(0, 0, L)$ Bragg peaks. To probe the possible development of in-plane FM below $T^{*}$, we measured the temperature dependence of these peaks.

Figure~\ref{00L_FM} shows the temperature dependence of the low-intensity $(0, 0, 2)$ nuclear reflection, chosen to maximize sensitivity to weak canting. Since magnetic scattering decreases with increasing $Q$, the strongest FM signal is expected at low-$L$ reflections. We observe no clear enhancement in the intensity of the $(0, 0, 2)$ Bragg peak below $T^{*} = 46$~K. Following the model procedure of Ref.~\citenum{Soh_2019}, we estimated the expected magnetic contribution to the $(0, 0, 2)$ peak for various canting angles $\alpha$; the results, shown in Fig.~\ref{00L_FM}(b), indicate that canting of $\gtrsim 3^\circ$ would be detectable in our measurements.

We also performed similar measurements and calculations for \cnmb\ crystals with $x = 0.05$ and $0.09$, where Na substitution raises $T^{*}$ to $\approx 75$~K and $\approx 100$~K, respectively \cite{Corasaniti_2019, Yang_2020}. As shown in Fig.~\ref{00L_FM_0p05} of Appendix~\ref{App:SCCNMB}, these results likewise indicate that substantial Mn moment canting is unlikely and, if present, must be $\lesssim 2^\circ$.

In Ref.~\citenum{Yang_2020}, a uniform canting angle of nearly $10^\circ$ was postulated for \cnmb\ with $x = 0$, $0.03$, and $0.05$. However, the absence of any discernible magnetic intensity enhancement at the $(0, 0, 2)$ reflection in our measurements indicates that uniform canting of the Mn moments, if present, must be $\lesssim 3^\circ$. Using an upper limit of $\alpha = 3^\circ$ and an ordered moment of $3.8\mu_\mathrm{B}$/Mn for \cmb\ \cite{Guo_PRB2014}, we estimate that the resulting in-plane FM moment due to canting must be less than $0.2~\mu_\mathrm{B}$/Mn.

Overall, our results indicate that significant spin canting capable of inducing in-plane ferromagnetism and breaking time-reversal symmetry, as required for a Weyl semimetallic state, is unlikely to occur in \cnmb. These intermediate results leave the origin of the anomalies observed in resistivity and optical conductivity unresolved. This raises two key questions that we address in the following sections: (i) What is the origin of the resistive anomaly in \cnmb? and (ii) Is there a detectable change in the magnetic or crystal structure associated with this anomaly?

\subsection{$L = 0.5$ Superlattice Reflections}
\label{SL_UPN}
\begin{figure}[]
\centering
\includegraphics[width=1.\columnwidth]{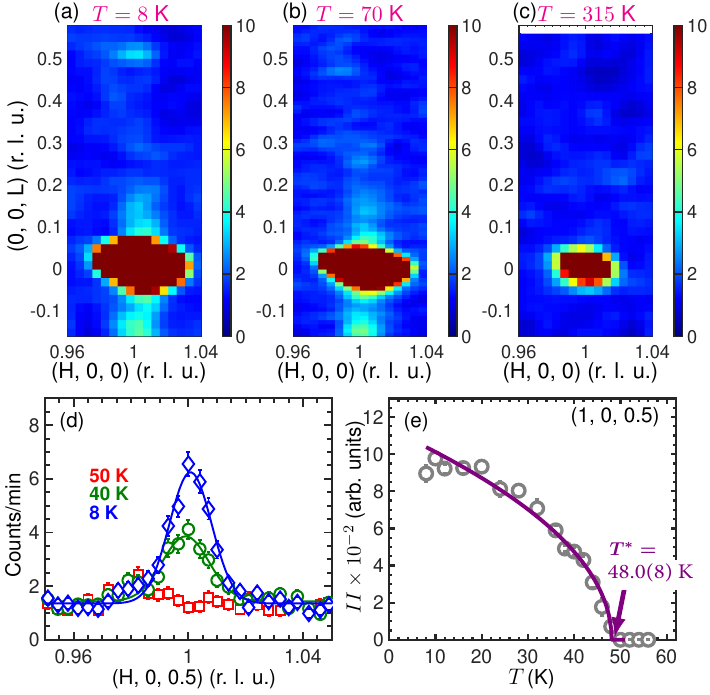}
\caption{Temperature dependence of the $(1, 0, 0.5)$ superlattice peak in \cmb. (a)–(c) Intensity contour plots from $(H, 0, L)$ mesh scans measured at three temperatures: $T < T^{*}$ (8~K), $T^{*} < T < T_\mathrm{N}$ (70~K), and $T > T_\mathrm{N}$ (315~K), respectively. (d) One-dimensional (1D) scans through $(1, 0, 0.5)$ along the $[H, 0, 0]$ direction at three representative temperatures. Solid lines are fits to a Gaussian lineshape. (e) Temperature dependence of the integrated intensity obtained by fitting 1D scans to Gaussian profiles, as shown in (d). The solid line is a fit to the power law $I(T) \propto (1 - T/T_\mathrm{N})^{2\beta}$, with $\beta = 0.23(1)$. }
\label{SPUP}
\end{figure}
To investigate possible changes in magnetism and lattice structure of \cmb\ across $T^{*}$, we mapped elastic scattering over a broad region of reciprocal space across several Brillouin zones using unpolarized neutrons, in search of possible superlattice or diffuse peaks. These measurements revealed a weak, commensurate superlattice peak at the $(1, 0, 0.5)$ position, which emerges at temperatures below $T^{*}$ \cite{TemperatureAccuracyNote}.

Figure~\ref{SPUP} shows contour plots of the measured neutron diffraction intensity including both the $(1, 0, 0.5)$ superlattice peak and the $(1, 0, 0)$ AFM Bragg peak at three different temperatures: $T < T^{*}$ (a), $T^{*} < T < T_\mathrm{N}$ (b), and $T > T_\mathrm{N}$ (c). In addition to the strong enhancement of the $(1, 0, 0)$ peak below the AFM ordering temperature, at $T = 8$K we observe weak superlattice intensity at $(1, 0, 0.5)$, which is absent at higher temperatures. One-dimensional (1D) scans along $(H, 0, 0.5)$ at three representative temperatures, shown in Fig.~\ref{SPUP}(d), confirm that the superlattice peak emerges below 50~K.

Figure~\ref{SPUP}(e) presents the temperature dependence of the integrated intensity of the $(1, 0, 0.5)$ superlattice peak, obtained by fitting the 1D scans such as shown in Fig.~\ref{SPUP}(d) to a Gaussian lineshape. The integrated intensity $II(T)$ evolves continuously, consistent with an order parameter in a second-order phase transition. A power-law fit yields a transition temperature of $T^{*} = 48.1(2)$~K, indicating that the observed superlattice peak is indeed associated with the resistivity anomaly at this temperature \cite{TemperatureAccuracyNote}.

Among a number of different Brillouin zones we surveyed, superlattice peaks were observed only at $(H, 0, L)$ positions with $H = 1$ and $L = 2n + \tfrac{1}{2}$. While the $H = 1$ condition, which corresponds to the C-type AFM ordering of in-plane Mn moments, could be taken as an indication of the involvement of magnetism in the observed superlattice structure, unpolarized neutron data alone cannot determine whether the peaks are magnetic, structural, or a combination of both. Indeed, as we show below, the superlattice peaks contain both structural and magnetic components. To clarify their origin, we carried out additional  x-ray diffraction and polarized neutron scattering measurements.

\subsection{X-Ray Diffraction Measurements: Tetragonal to Orthorhombic Transition}
\label{XRD}
\begin{figure}[]
\centering
\includegraphics[width=1.\columnwidth]{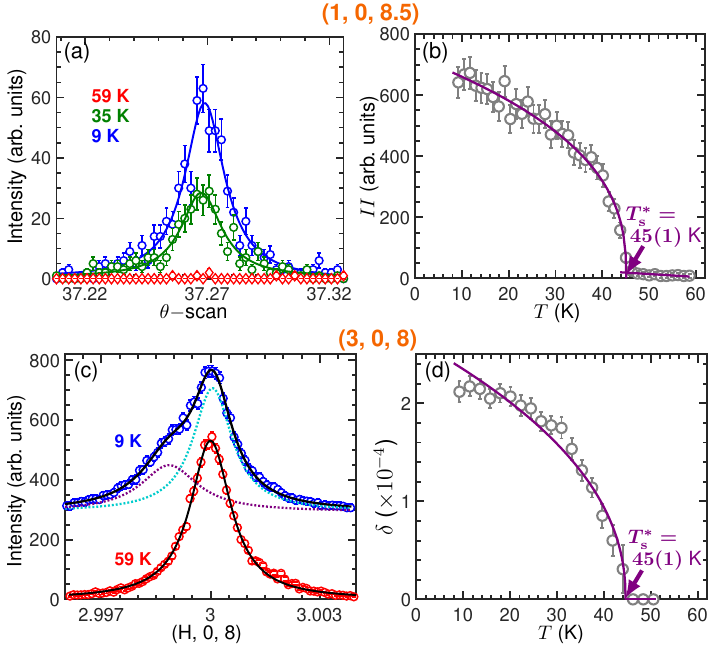}
\caption{X-ray diffraction measurements of a superlattice and a lattice Bragg peak in a \cmb\ single crystal. (a) Rocking curves ($\theta$-scans) at the $(1, 0, 8.5)$ superlattice peak at three different temperatures, showing its disappearance at $T = 59$~K~$> T^{*}$. Solid lines are fits to a Lorentzian lineshape. (b) Integrated intensity of the $(1, 0, 8.5)$ superlattice peak as a function of temperature. The solid line is an order-parameter fit with $\beta = 0.19(1)$. (c) $[H, 0, 0]$ scans through the $(3, 0, 8)$ lattice Bragg peak at two temperatures: $T > T^{*}$ (bottom) and $T < T^{*}$ (top); the data are vertically offset for clarity. The solid black lines represent a single-component Lorentzian fit at 59~K and a two-component fit at 9~K. Dotted colored lines show the Lorentzian components at 9~K, revealing orthorhombic splitting. (d) Orthorhombic lattice distortion, $\dfrac{a - b}{a + b}$, as a function of temperature. The solid line is an order-parameter fit with $\beta = 0.23(1)$. }
\label{xray}
\end{figure}

\begin{figure}[]
\centering
\includegraphics[width=1.\columnwidth]{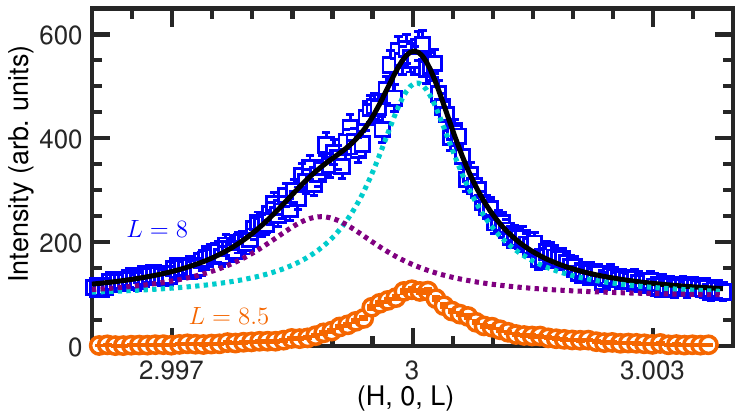}
\caption{X-ray diffraction measurements showing $H$-scans through the $(3, 0, 8.5)$ superlattice peak and the $(3, 0, 8)$ lattice peak in a \cmb\ single crystal at 9~K. Solid lines are fits to a single (superlattice) and two (lattice) Gaussian peaks. Error bars represent one standard deviation and where not visible are smaller than the symbol size.}%
\label{xray2}
\end{figure}

The results of x-ray diffraction measurements on a 100~mg single crystal of \cmb\ are presented in Figures~\ref{xray} and~\ref{xray2}. Superlattice peaks were measured in several Brillouin zones at wave vectors $(H, 0, 2n + \tfrac{1}{2})$, with $H = 0$, 1, 2, and 3. Consistent with the unpolarized neutron results, superlattice peaks at $L = 2n + \tfrac{1}{2}$ were observed for all measured integer values of $H$ except $H = 0$ [see Fig.~\ref{SS_H0andH2} in Appendix~\ref{App:SLPDBZ}]. The absence of superlattice peaks at $H = 0$ has important implications for the nature of the structural distortion. Specifically, it indicates that atomic displacements are perpendicular to the $c$-axis and therefore confined to the $ab$-plane.

Figure~\ref{xray}(a) shows the $(1, 0, 8.5)$ superlattice peak, which emerges below $T^{*}_\mathrm{s} = 45(1)$~K. This transition temperature was determined from an order-parameter fit to the temperature dependence of the peak’s integrated intensity, shown in Fig.~\ref{xray}(b). Additionally, the $H$-scan through the $(3, 0, 8)$ lattice Bragg reflection in Fig.~\ref{xray}(c) reveals peak splitting below $T^{*}_\mathrm{s}$, indicative of an orthorhombic lattice distortion. Together, the appearance of superlattice peaks in the x-ray measurements and the splitting of a lattice Bragg reflection provide clear evidence for a structural phase transition involving symmetry reduction from the high-temperature $P4/nmm$ phase.

The absence of splitting in the $(1, 0, 8.5)$ superlattice peak in Fig.~\ref{xray}(a) suggests that it originates from only one of the two symmetry-equivalent domains, meaning it appears at just one of the equivalent positions, $(H, 0, L)$ or $(0, K, L)$. This is reaffirmed by a similar observation in Fig.~\ref{xray2}, which shows $H$-scans through the same $(3, 0, 8)$ lattice peak as in Fig.~\ref{xray}(c) and the nearby $(3, 0, 8.5)$ superlattice peak. The absence of domain-related splitting in the superlattice peak supports the conclusion that this peak arises from a single domain orientation, in which atomic displacements are parallel to the Bragg wave vector of a superlattice. This further constrains the displacements to lie along either the $H$ or $K$ axis.

%

The distortion parameter, $\dfrac{a - b}{a + b}$, which quantifies the orthorhombic peak splitting in \cmb, falls within the same range as that observed in iron-based superconductors \cite{Nandi_2010}. This similarity suggests a possible connection to electronic nematicity, particularly notable given that \cmb\ is a semimetal. In this case, however, the nematic behavior would involve Dirac bands, consistent with the Peierls mechanism proposed by Hoffmann and co-workers \cite{TremelHoffman_JACS1987,PapoianHoffman_AngewChem2000}. The temperature dependence of the distortion parameter, shown in Fig.~\ref{xray}(d), with distortion increasing as temperature decreases, closely tracks the behavior of the superlattice peak intensity; both follow the temperature evolution of an order parameter in a second-order phase transition. A power-law fit to the temperature dependence of the distortion parameter yields a transition temperature $T^{*}_\mathrm{s} = 45(1)$~K, essentially the same as the value obtained from the superlattice peak intensity in Fig.~\ref{xray}(b).

It is clear from our unpolarized neutron and x-ray diffraction measurements that \cmb\ undergoes a structural phase transition at $T^{*}_\mathrm{s}$ from $P4/nmm$ to a lower-symmetry crystal structure \cite{TemperatureAccuracyNote}. Because the transition is second order, we used a group–subgroup relationship to infer the low-temperature symmetry for $T < T^{*}_\mathrm{s}$, employing the ISODISTORT \cite{Campbell_2006} and ISOSUBGROUP \cite{Stokes_2016} tools within the ISOTROPY software suite \cite{Stokes_iso}. Further details are given in Appendix~\ref{App:ISODISTORT}.

We find that only the orthorhombic $Pcmn$ subgroup ($\#62$, $Pnma$ in standard setting) permits displacement modes that produce superlattice peaks consistent with our experimental observations. For comparison, no such mode or combination of modes exists within the orthorhombic space group $Cmca$. Furthermore, the symmetry lowering from $P4/nmm$ to $Pcmn$ also permits splitting of the $(H, 0, L)$ Bragg peaks, as observed in Figs.~\ref{xray}(c) and \ref{xray2}, due to the formation of twin domains.

$Pcmn$ is a non-standard setting of orthorhombic space group 62, whose standard setting, $Pnma$, is commonly used to describe other 112 compounds. We adopt the $Pcmn$ setting to maintain consistent axis orientation and labeling with respect to the high-temperature tetragonal $P4/nmm$ unit cell. The basis vectors of the $Pcmn$ subgroup which realizes the $Z^{-}_5$ irreducible representation (IR) are (0, –1, 0), (1, 0, 0), and (0, 0, 2) relative to the parent $P4/nmm$ lattice. These correspond to $\ba_\mathrm{orth} = -\bb_\mathrm{tet}$, $\bb_\mathrm{orth} = \ba_\mathrm{tet}$, and $\bc_\mathrm{orth} = 2\bc_\mathrm{tet}$ for the $Pcmn$ superlattice unit cell. For the conjugate subgroup which realizes the $Z^{+}_5$ IR, the basis vectors are (1, 0, 0), (0, 1, 0), and (0, 0, 2), yielding $\ba_\mathrm{orth} = \ba_\mathrm{tet}$, $\bb_\mathrm{orth} = \bb_\mathrm{tet}$, and $\bc_\mathrm{orth} = 2\bc_\mathrm{tet}$. Thus, in both cases, the long $c$-axis of $Pcmn$ remains aligned with that of $P4/nmm$ [Fig.~\ref{CMS}], in contrast to the $Pnma$ setting, where the $c$-axis of $P4/nmm$ becomes the $a$-axis of $Pnma$ \cite{IntTablesCrystA_6thEd2016}. 

The distinction between the superlattices associated with the $Z^{-}_5$ and $Z^{+}_5$ IRs cannot be resolved from the x-ray data and is addressed below in the neutron diffraction refinement sections. Figure~\ref{CMS}(a) illustrates the $Z^{-}_5$ low-temperature superlattice, as determined from our refinement of the neutron diffraction data.

To summarize, our findings show that \cmb\ undergoes a second-order structural phase transition from the high-temperature tetragonal $P4/nmm$ structure to an orthorhombic $Pcmn$ lattice below $T^{*}_\mathrm{s} \approx 46$~K, coinciding with the resistive anomaly \cite{TemperatureAccuracyNote}. The absence of superlattice intensity at $(0, 0, L + 0.5)$ positions indicates that the atomic displacements associated with the symmetry-lowering transition are confined to the $ab$-plane, consistent with layer-shearing distortions such as those observed in YbMnSb$_2$ \cite{Bhoi_PRB2025, Hu_unpublished2025}. This is further confirmed by our symmetry analysis. The absence of twinning of the $(1, 0, L+0.5)$ superlattice reflections indicates collinear displacements along the $H$ or $K$ direction, corresponding to the two orthorhombic twin domains. However, x-ray and unpolarized neutron measurements alone cannot determine whether the observed superlattice peaks include a magnetic contribution or are otherwise linked to magnetism. \\

\subsection{Polarized Neutron Diffraction Evidence for Structural and Magnetic Superlattice Order}
\label{PND}

\begin{figure}[]
\centering
\includegraphics[width=1.\columnwidth]{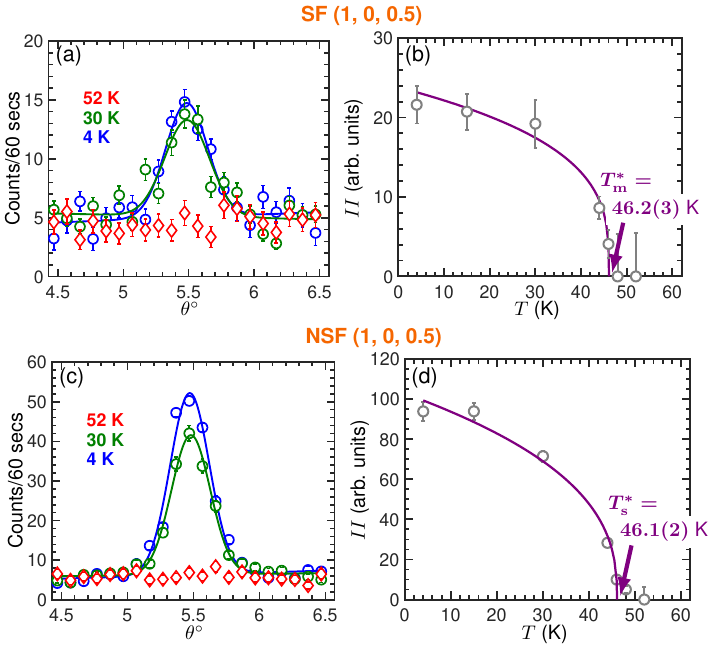}
\caption{Temperature dependence of the $(1, 0, 0.5)$ superlattice peak from polarized neutron diffraction measurements on a \cmb\ single crystal: spin-flip (SF) channel [(a) and (b)] and non-spin-flip (NSF) channel [(c) and (d)]. (a) and (c) Rocking curves ($\theta$-scans) of the $(1, 0, 0.5)$ superlattice peak at three characteristic temperatures for SF and NSF, respectively. (b) and (d) Temperature dependence of the integrated intensity ($II$), obtained by fitting the 1D scans in (a) and (c) with Gaussian lineshapes (solid lines). The solid lines in (b) and (d) are fits to a power law, $II \propto (1 - T/T^{*})^{2\beta}$, yielding $T^{*}_\mathrm{m} \approx T^{*}_\mathrm{s} \approx 46$~K and critical exponents $\beta = 0.16(2)$ and $0.19(3)$ for the SF and NSF channels, respectively. }
\label{SPP}
\end{figure}

To test whether the superlattice peaks observed below $T^{*}$ have a magnetic component and to clarify its relation to the structural transition identified by unpolarized neutron and x-ray diffraction, we carried out polarized neutron diffraction experiments. Measurements were performed on the HB-1 triple-axis spectrometer at HFIR with neutron polarization aligned along the momentum transfer, $\mathbf{P} \parallel \mathbf{Q}$. This configuration directly separates magnetic and nuclear scattering into spin-flip (SF) and non-spin-flip (NSF) channels, respectively \cite{Boothroyd_book_2017}. All measured intensities were corrected for the flipping ratio, FR = 16, determined from nuclear Bragg peak intensities measured in both channels \cite{Zaliznyak_2017}.

Results of the polarized neutron measurements of the $(1, 0, 0.5)$ superlattice peak as a function of temperature are shown in Fig.~\ref{SPP}. For a purely structural (nuclear) reflection, no intensity is expected in the SF channel. In contrast, as illustrated by the rocking curves in Fig.~\ref{SPP}(a) and (c), the superlattice peak emerges below the resistive anomaly temperature, $T^{*} \approx 46$~K, with comparable intensity (within a factor of $\sim 3$) in both the NSF and SF channels.

By fitting the temperature dependence of the superlattice intensity to a critical-type power law, we obtain transition temperatures of $T^{*}_\mathrm{m} = 46.2(3)$K for the magnetic (SF) channel [Fig.~\ref{SPP}(b)] and $T^{*}_\mathrm{s} = 46.1(2)$K for the structural (NSF) channel [Fig.~\ref{SPP}(d)]. The two values are indistinguishable within experimental uncertainty, indicating a coupled structural and magnetic transition.

The critical exponents extracted from the power-law fits, $\beta = 0.16(2)$ for the SF channel and $\beta = 0.19(3)$ for the NSF channel, are essentially identical within uncertainty. Both are significantly smaller than those expected for three-dimensional models ($\beta = 0.365$ for the 3D Heisenberg and $\beta \approx 0.326$ for the 3D Ising), indicating reduced dimensionality of the observed magneto-structural transition. Notably, these values are {not far from} the two-dimensional (2D) Ising result, $\beta = 1/8$, consistent with expectations for a tetragonal-to-orthorhombic transition in a layered crystal, where the symmetry breaking maps onto an effective 2D Ising model \cite{ZacharZaliznyak_PRL2003, Zaliznyak_PRB2012}.

\subsection{Symmetry of the $Pcmn$ Distortion}
\label{Pcmn_symmetry}
As discussed in Section~\ref{XRD} and Appendix~\ref{App:ISODISTORT}, only the orthorhombic $Pcmn$ subgroup of the high-temperature tetragonal $P4/nmm$ structure permits atomic displacements consistent with both the observed superlattice peaks and a continuous structural transition. This subgroup contains two irreducible representations (IRs) of the parent $P4/nmm$ space group, $Z^{-}_5$ and $Z^{+}_5$, which differ by an origin shift of the $Pcmn$ unit cell by (0, 0, 0) and (0, 0, 1/2), respectively, relative to the $P4/nmm$ lattice.

\begin{figure}[h!]
\centering
\includegraphics[width=1.0\columnwidth]{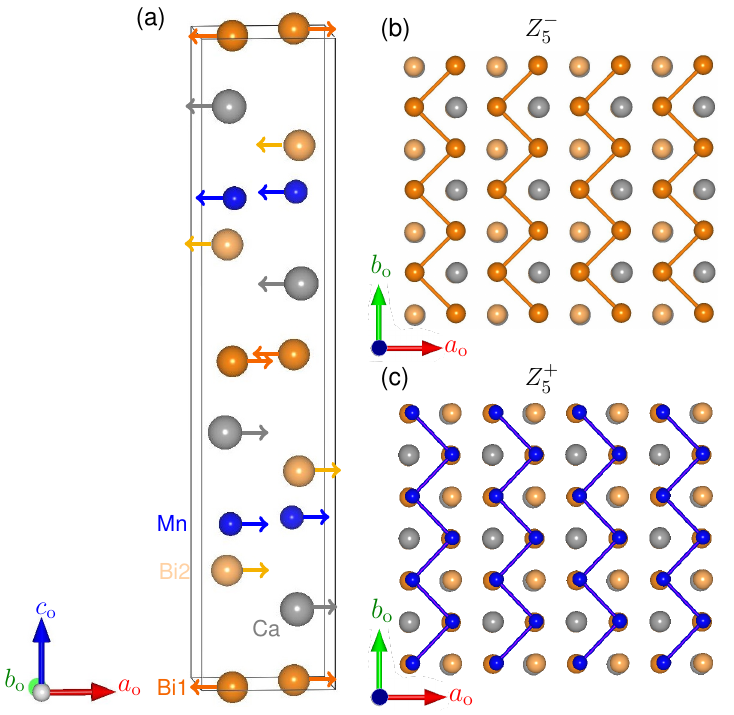}
\caption{(a) Orthorhombic $Pcmn$ crystal structure of \cmb, illustrating $Z^{-}_5$ in-plane atomic displacements along the $a_\mathrm{o}$ direction, which is equivalent to the $b_\mathrm{tet}$ direction. Arrows indicate the displacements of each atom along the $a$-axis of $Pcmn$ or $P4/nmm$ (equivalent to displacements along the $c$-axis in the standard $Pnma$ setting \cite{Bhoi_PRB2025}). (b) and (c) Bi–Bi and Mn–Mn bond-wave orders corresponding to the irreducible representations $Z^{-}_5$ and $Z^{+}_5$, respectively. }
\label{Fig:IPD}
\end{figure}

In the $Z^{-}_5$ case, Bi1 atoms in the square-net layers at $z = 0$ undergo anti-phase displacements, producing bond disproportionation and zig-zag Bi chains staggered between adjacent Bi1 layers, leading to lattice doubling. {Thus, the refined displacements directly produce alternating in-plane Bi--Bi bonds, forming the zigzag BOW. This bond modulation is the structural ingredient of the Peierls mechanism that reconstructs the Bi-derived Dirac bands, opening a gap along the dimerization direction while preserving the Dirac nodes along the perpendicular zigzag direction [Fig.~\ref{CMS}(e)].} For the intra-unit-cell Bi2, Mn, and Ca layers, in-phase, shear-type displacements are allowed [Fig.~\ref{Fig:IPD}(a),(b)]. This structure corresponds to the Peierls-type distortion of Bi layers proposed by Hoffmann and co-workers \cite{TremelHoffman_JACS1987,PapoianHoffman_AngewChem2000}. Conversely, in the $Z^{+}_5$ case, zig-zag chains form in the Mn layers at $z = 0$ and $z = 0.5$ of the doubled $Pcmn$ unit cell, while other intra-cell layers undergo shear-type displacements. The resulting in-plane Mn-Mn bond order [Fig.~\ref{Fig:IPD}(c)] is similar to the bond-order wave associated with ferro-orbital ordering observed in the Fe layers of FeTe \cite{Fobes_PRL2014}. 

For the magnetic Rietveld refinement, maximal magnetic symmetries consistent with the $Pcmn$ ($Pnma$) crystal symmetry and propagation vector $\mathbf{k} = (1, 0, 0)$ were generated using the magnetic symmetry analysis tools of the Bilbao Crystallographic Server \cite{Bilbao_I,Bilbao_II,Bilbao_III,Mato_bilbao_maxmag}. Among the tested configurations, the $Pc^\prime m^\prime n^\prime$ ($Pn^\prime m^\prime a^\prime$; $\#62.449$) magnetic symmetry, defined by the operations
$(x, 1/4, z | M_x, 0, M_z)$,
$(-x+1/2, \textcolor{black}{1/4}, z+1/2 |  -M_x, 0, M_z)$,
$(-x, 3/4, -z |  -M_x, 0, -M_z)$,
$(x+1/2, \textcolor{black}{3/4}, -z+1/2 |  M_x, 0, -M_z)$,
provided the best fit to the data, as discussed below. This symmetry corresponds to a C-type magnetic structure that allows canting of Mn magnetic moments with $M_x$ component staggered from layer to layer.

Table~\ref{Table:Mn_pos_FHKL} lists the positions of the Mn ions and their magnetic moments in the unit cell of the $Pc^\prime m^\prime n^\prime$ structure for both the $Z^{-}_5$ and $Z^{+}_5$ IRs. The table also gives the magnetic contributions to the structure factors of the main-lattice ($Pcmn$ $L$ even) and superlattice ($Pcmn$ $L$ odd) Bragg peaks, arising from the two twinned domains, $(H,0,L)$ and $(0,H,L)$, that contribute to the intensity at each position.

For the $Z^{-}_5$ IR, the superlattice magnetic intensity $\sim M_z^2 \delta^2$ arises from the displacements $\delta$ of the ordered Mn moments and is insensitive to moment canting, because contributions from a uniform ferromagnetic (FM) component of the two in-plane Mn moments in the unit cell cancel. Consequently, the potential in-plane FM component $M_x$ cannot be determined from the magnetic superlattice intensity measured by polarized neutron diffraction.

For the $Z^{+}_5$ IR, the canting is uniform and there is no net in-plane FM moment. The $M_x$ components of the two in-plane Mn moments in the unit cell are oppositely directed, producing magnetic intensity $\sim M_x^2$. In this case, however, there is no contribution from the displacements $\delta$ of the ordered $M_z$ moments ($\sim M_z^2$), so the superlattice magnetic intensity vanishes without canting and is therefore expected to be very small, in contrast to our observations in Fig.~\ref{SPP}.

\newcolumntype{C}{>{$}c<{$}}
\begin{table}[t!]
\centering
\renewcommand{\arraystretch}{1.25}
\caption{\label{Table:Mn_pos_FHKL}
Magnetic moments and positions of the four Mn ions in the $Pcmn$ unit cell for the $Z^{-}_5$ and $Z^{+}_5$ IRs. The lower two rows show the dependence of the squared magnetic structure factors, $|F_M(H,0,L)|^2$ and $|F_M(0,H,L)|^2$, on the ordered moment, representing the contributions of the two orthorhombic domains to the magnetic intensity at $(H,0,L)$ for odd $H$ and odd $L$. Odd $L$ corresponds to $L = 0.5$ in the parent $P4/nmm$ lattice, and thus to superlattice Bragg peaks. 
}
\begin{tabular}{@{} l c | C C C | C C C @{}}
\toprule
 & \multirow{2}{*}{Moment} & \multicolumn{3}{c|}{$Z^{-}_5$ IR} & \multicolumn{3}{c}{$Z^{+}_5$ IR} \\
\cmidrule(lr){3-5}\cmidrule(l){6-8}
Mn site &  & x & y & z & x & y & z \\
\midrule
Mn1 & $(M_x,0,M_z)$   & \tfrac{1}{4}+\delta & \tfrac{1}{4} & \tfrac{3}{4} & \tfrac{1}{4}+\delta & \tfrac{1}{4} & 0 \\
Mn2 & $(M_x,0,-M_z)$  & \tfrac{3}{4}+\delta & \tfrac{3}{4} & \tfrac{3}{4} & \tfrac{3}{4}+\delta & \tfrac{3}{4} & \tfrac{1}{2} \\
Mn3 & $(-M_x,0,-M_z)$ & \tfrac{3}{4}-\delta & \tfrac{3}{4} & \tfrac{1}{4} & \tfrac{3}{4}-\delta & \tfrac{3}{4} & 0 \\
Mn4 & $(-M_x,0,M_z)$  & \tfrac{1}{4}-\delta & \tfrac{1}{4} & \tfrac{1}{4} & \tfrac{1}{4}-\delta & \tfrac{1}{4} & \tfrac{1}{2} \\
\midrule
\multicolumn{2}{l|}{$|F_M(H,0,L)|^2 \sim$} & \multicolumn{3}{l|}{$M_z^2 \sin^2(2\pi H\delta)$} & \multicolumn{3}{l}{$M_x^2 \cos^2(2\pi H\delta)$} \\
\multicolumn{2}{l|}{$|F_M(0,H,L)|^2 \sim$} & \multicolumn{3}{l|}{0} & \multicolumn{3}{l}{$M_x^2$} \\
\bottomrule
\end{tabular}
\end{table}

\subsection{Refinement of Crystal and Magnetic Structures by Polarized and Unpolarized Neutron Diffraction}
\label{PTAX_and_HB3A}

Following the symmetry analysis, two space groups were considered for the crystal structure refinement at 5~K: tetragonal $P4/nmm$ and orthorhombic $Pcmn$. The $Pcmn$ subgroup imposes a constraint that forbids atomic displacements of Mn (for IR $Z^{+}_5$) or Bi1 (for IR $Z^{-}_5$) atoms along the $c$-axis of $P4/nmm$, consistent with the observed absence of $(0,0,L+0.5)$ superlattice reflections. The $z$-coordinates of Mn and Bi1 atoms were therefore kept fixed to their $P4/nmm$ nominal values in our refinements, thus limiting the atomic displacements to the $ab$-plane. Because only a limited number of Bragg peaks were measured, isotropic atomic displacement parameters (ADPs) were used in the refinements.

To determine the low-temperature crystal and magnetic structure at $T = 5$~K ($< T^{*}$), we first performed a limited polarized-neutron refinement by measuring a set of 14 nuclear, magnetic, and superlattice Bragg peaks across several Brillouin zones in both NSF and SF channels on the HB-1 TAS. Rietveld refinements were then carried out with the FullProf suite \cite{Rodriguez_fullprof}, using integrated intensities extracted from rocking scans. Further details are provided in Appendices~\ref{App:TAS_refinement} and \ref{App:PND_refinement}.

\begin{figure}[t!]
\centering
\includegraphics[width=1.0\columnwidth]{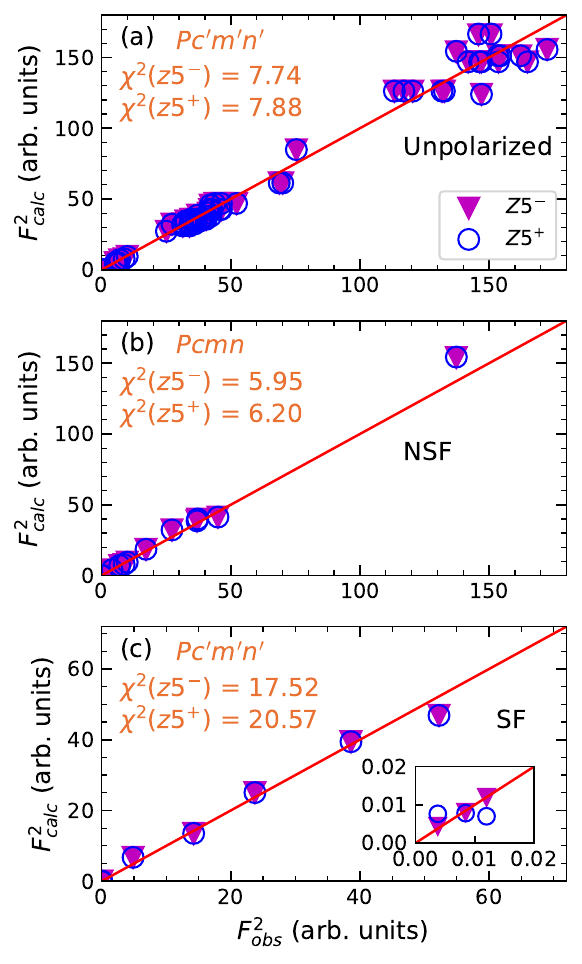}
\caption{Summary of the combined refinements of \cmb\ at 5~K, using Bragg peak intensities from (a) unpolarized measurements, (b) NSF channel, and (c) SF channel. The plots compare calculated and observed intensities, expressed as squared structure factors in arbitrary units. Magenta triangles correspond to $Z^{-}_5$ and blue open circles to $Z^{+}_5$. The inset highlights the low-intensity superlattice peaks. \label{Fig:combined_refinement}}
\end{figure}

Our polarized TAS refinements at 5~K clearly rule out the tetragonal $P4/nmm$ structure, establishing that the low-temperature phase is orthorhombic $Pcmn$. They also consistently yield slightly lower $\chi^2$ values for the $Z^{-}_5$ model, corresponding to the Peierls-type formation of Bi zigzag chains predicted by Hoffmann and collaborators. However, the atomic positions and other structural parameters cannot be reliably determined because only a limited number of Bragg peaks were measured with polarized neutrons.

Consequently, we carried out supplementary refinements with unpolarized neutrons using the HB-3A DEMAND diffractometer. Details are provided in Appendix~\ref{App:HB3A_refinement}. The $L = 2n + \tfrac{1}{2}$ superlattice peaks (indexed in tetragonal $P4/nmm$) are too weak to be detected in this measurement. To improve the refinement, we supplemented the HB-3A dataset with three superlattice peaks, $(1,0,0.5)$, $(1,0,1.5)$, and $(1,0,2.5)$, observed on the HB-1 triple-axis spectrometer. Their intensities were re-scaled using the average ratio of the four $(H,0,L)$ main lattice Bragg peaks ($H = 1$, $L = 0,2,4,6$ in $Pcmn$ notation) measured on both instruments. We then refined this combined dataset together with the NSF and SF Bragg intensities measured with polarized neutrons on HB-1, using a single set of lattice and magnetic structure parameters for each of the two models corresponding to the irreducible representations $Z^{-}_5$ and $Z^{+}_5$ of $Pcmn$. The refinement results are presented in Fig.~\ref{Fig:combined_refinement} and Table~\ref{Table:combined_refinement}.

\begin{table}[h!]
\caption{\textbf{HB3A \& TAS combined refinements.} Crystal and magnetic structure parameters of \cmb\ at 5~K obtained from Rietveld refinement of integrated Bragg intensities measured on HB-3A and HB-1. Peak intensities from HB-1 were re-scaled to HB-3A as described in text. Spin canting is not resolvable in the $Z^{-}_5$ model. Lattice parameters: $a = 4.456$~\AA, $b = 4.459$~\AA, $c = 21.158$~\AA. \label{Table:combined_refinement}}
\begin{ruledtabular}
\begin{tabular}{ccccc}
T = 5~K & \multicolumn{2}{c}{$Pc'm'n'$} $Z^{-}_5$& \multicolumn{2}{c}{}\\
\hline
Atom & $x$ & $y$ & $z$ &$B_\mathrm{iso}$ (\AA$^2$)\\
\hline
Ca  & 0.7705(2) & 0.2500 & 0.6178(2) & 0.243(47)\\
Mn  & 0.2534(1) & 0.2500 & 0.7500 & 0.542(43)\\
Bi1 & \textcolor{red}{0.2465(1)} & 0.2500 & 0.5000 & 0.350(22)\\
Bi2 & 0.7590(1) & 0.2500 & 0.8303(1) & 0.435(26)\\
\hline
\multicolumn{5}{c}{$\mu_{\mathrm{Mn}}(x) = N/A$, $\mu_{\mathrm{Mn}}(z) = 3.97(2) \mu_B$}\\
\end{tabular}
\end{ruledtabular}
\begin{ruledtabular}
\begin{tabular}{ccccc}
T = 5~K & \multicolumn{2}{c}{$Pc'm'n'$} $Z^{+}_5$& \multicolumn{2}{c}{}\\
\hline
Atom & $x$ & $y$ & $z$ &$B_\mathrm{iso}$ (\AA$^2$)\\
\hline
Ca  & 0.7646(3) & 0.2500 & 0.8680(2) & 0.412(47)\\
Mn  & \textcolor{red}{0.2539(2)} & 0.2500 & 0.0000 & 0.554(46)\\
Bi1 & \textcolor{red}{0.2420(126)} & 0.2500 & 0.2500 & 0.313(156)\\
Bi2 & 0.7331(2) & 0.2500 & 0.0803(1) & 0.284(27)\\
\hline
\multicolumn{5}{c}{$\mu_{\mathrm{Mn}}(x) = 0.062(3) \mu_B$, $\mu_{\mathrm{Mn}}(z) = 3.97(2) \mu_B$}\\
\end{tabular}
\end{ruledtabular}
\end{table}

The combined refinements enable determination of the isotropic ADPs ($B_\mathrm{iso}$) and other structural and magnetic parameters listed in Table~\ref{Table:combined_refinement}, and consistently indicate a preference (lower $\chi^2$) for the $Z^{-}_5$ model. In the $Z^{-}_5$ IR, the displacements correspond to oppositely directed shearing-type shifts of atoms in the upper and lower halves of the unit cell, with the central Bi1 layer exhibiting anti-phase displacements that produce a bond-order wave forming zigzag chains, consistent with the Peierls-type distortion \cite{TremelHoffman_JACS1987,PapoianHoffman_AngewChem2000} and similar to that reported for YbMnSb$_2$~\citenum{Bhoi_PRB2025}. In the $Z^{+}_5$ IR, a comparable displacement pattern occurs, but with a zigzag Mn–Mn bond order in the Mn square lattice (Fig.~\ref{Fig:IPD}).

The refinements, however, still exhibit several nearly equal-depth $\chi^2$ minima within $x \in [0.2465, 0.258]$ for the Bi1 displacement along the $a$-axis in $Z^{-}_5$, and an additional broad $\chi^2$ minimum near $x = 0.27$ for the Mn displacement in $Z^{+}_5$, similar to the NSF-only polarized TAS lattice refinements (Appendix~\ref{App:TAS_refinement}). This indicates that the corresponding parameters remain poorly constrained, although the minimum and maximum displacements are reliably determined. Additionally, in the $Z^{+}_5$ model, the shearing displacement of the Bi1 layer is also ill-defined, as reflected by its unrealistically large uncertainty. This suggests the structural deficiency of the $Z^{+}_5$ model, where the magnetic component of the superlattice reflections arises solely from in-plane-staggered spin canting (Table~\ref{Table:Mn_pos_FHKL}) that also contributes to other lattice reflections. In Table~\ref{Table:combined_refinement}, poorly constrained parameters are shown in red; the small error bars on the displacements represent the nominal refinement uncertainty within the first $\chi^2$ minimum.

From the magnetic refinements, we obtain the ordered magnetic moment per Mn ion in \cmb\ as $\mu_{\mathrm{Mn}}(z) = 3.68(5) \mu_B$ at 50~K and $\mu_{\mathrm{Mn}} \approx 4 \mu_B$ at 5~K [$\mu_{\mathrm{Mn}}(z) = 3.97(2),\mu_B$ for both $Z^{-}_5$ and  $Z^{+}_5$, with a small staggered in-plane component $\mu_{\mathrm{Mn}}(x) = 0.062(3) \mu_B$, corresponding to a canting angle of about $1.2^\circ$ for $Z^{+}_5$]. Here, $\mu_{\mathrm{Mn}}(z)$ corresponds to the component of the ordered moment along the $c$-axis of the tetragonal $P4/nmm$ lattice, while $\mu_{\mathrm{Mn}}(x)$ represents the in-plane component arising from spin canting.

\section{DFT Analysis of Crystal Structure and Electronic Properties}
\label{DFT}

\begin{table*}[t!]
\caption{\label{Table:DFT}%
Lattice parameters and energy differences per formula unit for orthorhombic and tetragonal phases of \cmb\ from DFT (GGA functional, C-type AFM order). Similar results obtained with SCAN functional. Experimental data: $^{a}$ Springer Materials; $^{b}$ this work. }
\begin{ruledtabular}
\begin{tabular}{lcccccc}
Phase & $a$ (\AA) & $b$ (\AA) & $c$ (\AA) & $d_{\textrm{Bi1--Bi1}}$ (\AA) & $d_{\textrm{Mn--Mn}}$ (\AA) & $\Delta E$ (meV)\\
\hline
$P4/nmm$ (exp)$^{a}$        & 4.450 & 4.450 & 11.082 & 3.147          & 3.147          & -- \\
$P4/nmm$ (DFT-relaxed)      & 4.526 & 4.526 & 10.838 & 3.197          & 3.197          & 0  \\
$Pcmn$ $Z_5^-$ (exp)$^{b}$  & 4.483 & 4.481 & 21.968 & 3.154, 3.184   & 3.169          & 29 \\
$Pcmn$ $Z_5^+$ (exp)$^{b}$  & 4.483 & 4.481 & 21.968 & 3.169          & 3.151, 3.188   & 35 \\
$Pcmn$ (DFT-relaxed)        & 4.525 & 4.525 & 21.680 & 3.197, 3.203   & 3.200          & 0  \\
\end{tabular}
\end{ruledtabular}
\end{table*}

To gain further insight into the relationship between the electronic and structural properties of \cmb, and to complement our experimental findings, we performed density functional theory (DFT) calculations aimed at characterizing the electronic origin of the structural distortion, evaluating the relative stability of competing lattice distortion modes, and probing possible Fermi surface reconstruction across the transition at $T^{*}$. The results of these analyses are presented below. Additional details of the computational methods and the resulting total energy and electronic structure analysis are provided in Appendix~\ref{DFT_appendix}.

\subsection{Total Energy of Competing Lattice Distortions}
\label{DFT_Total_Energy}

To evaluate the relative stability of different lattice distortion modes in \cmb, we performed total energy calculations for several candidate structures, including the high-symmetry tetragonal $P4/nmm$ phase and symmetry-lowered orthorhombic configurations corresponding to the irreducible representations $Z^{-}_5$ and $Z^{+}_5$. These distorted structures were generated by introducing displacement patterns obtained from ISODISTORT analysis and were subsequently relaxed under symmetry constraints. The resulting total energies were then compared to identify the energetically preferred distortion mode and to estimate the driving force behind the structural phase transition at $T^*$. All calculations were carried out at zero temperature and fixed experimental volume, unless otherwise specified, similar to the analysis of EuZnSb$_2$ \cite{Aryal_PRB2022}.

Table~\ref{Table:DFT} summarizes the total energies and lattice parameters for the tetragonal and various orthorhombic phases, based on both experimentally refined and DFT-relaxed structures. When comparing the experimentally refined orthorhombic structures corresponding to the $Z_5^-$ (dimerization in the Bi layer) and $Z_5^+$ (dimerization in the Mn layer) irreducible representations, the $Z_5^-$ phase is found to be slightly lower in energy. However, in full lattice relaxation calculations initiated with varying dimerization amplitudes, both the $Z_5^-$ and $Z_5^+$ configurations ultimately relax back to the high-symmetry tetragonal phase.

We further performed total energy calculations exploring the space of shear and dimerization amplitudes. Figure~\ref{Fig:DFT_Fig1S_energy}(a) in Appendix~\ref{DFT_appendix} shows the energy cost for varying dimerization in the Mn and Bi layers, without introducing shear distortion. While dimerization in the Bi square layer is slightly more favorable energetically than in the Mn layer, consistent with our earlier result that the $Z_5^-$ structure is slightly lower in energy than the $Z_5^+$ structure (Table~\ref{Table:DFT}), both configurations correspond to higher-energy states in our calculations. Figures~\ref{Fig:DFT_Fig1S_energy}(b) and \ref{Fig:DFT_Fig1S_energy}(c) in Appendix~\ref{DFT_appendix} present the total energy landscape as a function of both dimerization and shear distortions for the $Z_5^-$ and $Z_5^+$ cases, respectively. In our DFT analysis, the minimum total energy always corresponds to the undistorted tetragonal phase, indicating that the energy gain associated with the distortion is very small and within the uncertainty of the DFT calculations. This result is not unexpected, given that the transition temperature $T^*$ is only $\approx 46$~K, corresponding to a thermal energy scale of $\approx 4$~meV.

\subsection{Effect of Lattice Distortion on Electronic Band Structure}
\label{Dispersion}

\begin{figure*}[]
\centering
\includegraphics[width=0.95\textwidth]{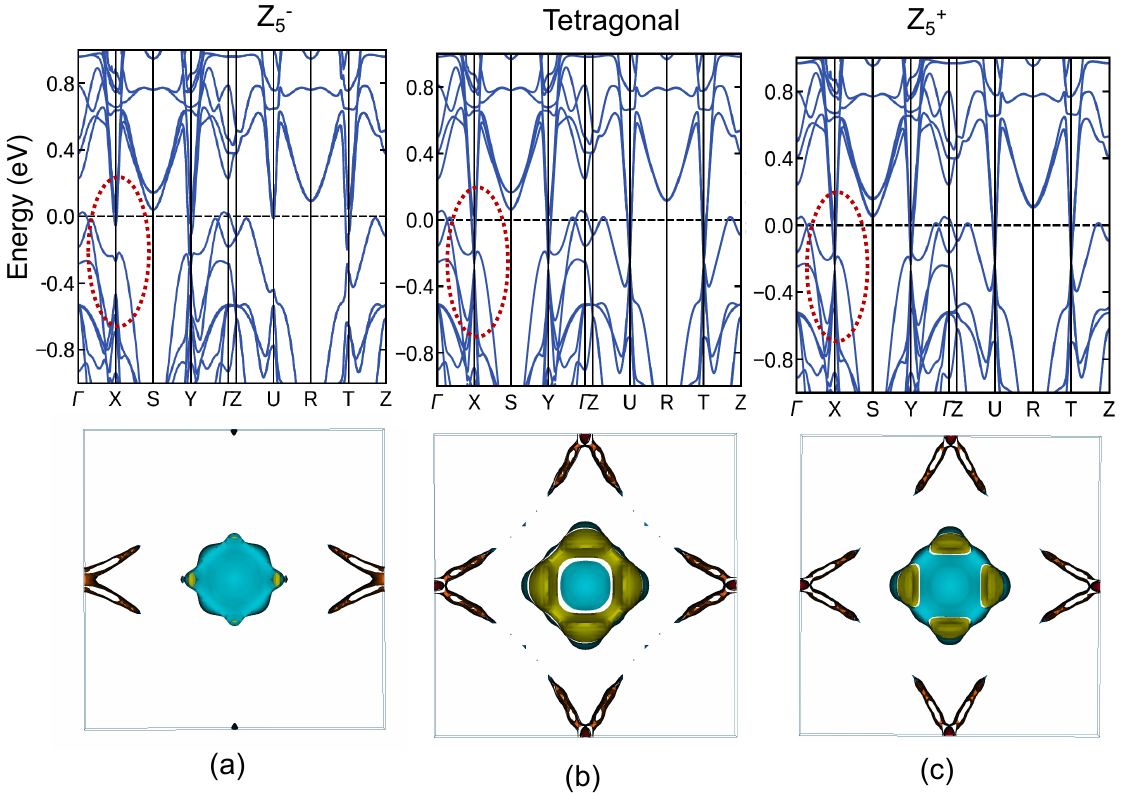}
\caption{\label{fig:bandFScomparison}
Band structure and Fermi surface (FS) plots for (a) the Bi-dimerized $Z_5^-$ phase, (b) the undimerized tetragonal phase, and (c) the Mn-dimerized $Z_5^+$ phase. Major differences in the band structure near the Fermi level are highlighted by red-dotted circles. }
\end{figure*}

We first calculated the electronic band structure and density of states (DOS) for the fully relaxed tetragonal phase of CaMnBi$_2$, without including spin–orbit coupling (SOC), as shown in Fig.~\ref{Fig:DFT_Fig2S_bands} of Appendix~\ref{DFT_appendix}. Consistent with previous studies, the Bi 5\textit{p} states near the Fermi level, originating from the square-net Bi atoms, give rise to Dirac nodal points {at $\approx -0.3$~eV}.

To understand the impact of BOW dimerization on the electronic structure, we then compared the DOS for the dimerized phases [Fig.~\ref{Fig:DFT_Fig2S_bands}(b) in Appendix~\ref{DFT_appendix}]. The DOS profiles for the experimentally refined $Z_5^-$ and $Z_5^+$ distortions are nearly identical, as expected given the small DOS of Dirac nodes and very small experimental dimerization amplitude ($\sim$10$^{-3}$\AA). Upon closer inspection, however, the $Z_5^-$ phase exhibits a slightly reduced DOS near the Fermi level compared to $Z_5^+$. Since this difference lies within the typical uncertainty of DFT calculations, we artificially increased the dimerization strength to 0.015\AA\ in the relaxed structure to further examine whether the Bi-dimerized phase indeed yields a lower DOS and opens a charge density wave (CDW) gap.

With enhanced dimerization, the Bi-dimerized $Z_5^-$ phase shows a clear reduction in the DOS near the Fermi level, due to a redistribution of spectral weight within the Bi $p_x$–$p_y$ orbitals, as seen in the lower panel of Fig.~\ref{Fig:DFT_Fig2S_bands}(b) in Appendix~\ref{DFT_appendix}. This is consistent with the fact that the states near the Fermi level in the undistorted tetragonal phase are predominantly contributed by Bi 5\textit{p} orbitals from the square-net layer {and with the Peierls-type band structure reconstruction illustrated in Fig.~\ref{CMS}(d),(e)}.

To further investigate the impact of dimerization on the electronic structure and Fermi surface topology, we calculated the band dispersion and Fermi surfaces for the Bi- and Mn-dimerized phases with a dimerization strength of 0.015~\AA. The results, shown in Fig.~\ref{fig:bandFScomparison}, reveal marked changes in the Bi-dominated bands near the Fermi level for the $Z_5^-$ phase, similar to those reported for YbMnSb$_2$ \cite{Bhoi_PRB2025}. {These changes are consistent with the predicted opening of a Peierls gap~\cite{TremelHoffman_JACS1987,PapoianHoffman_AngewChem2000} along the dimerization direction (X point), while Dirac nodes are preserved along the perpendicular zigzag direction (Y point), as illustrated in Fig.~\ref{CMS}(e)}. A partial gapping of the Fermi surface is consistent with the resistive anomaly associated with the dimerization transition. In contrast, only minor changes in the band structure and Fermi surface are observed in the $Z_5^+$ phase.

\section{Discussion}
\label{Sec:Discussion}

The relevance of weak ferromagnetism in $A$Mn$X_2$ compounds has been controversial. In SrMnSb$_2$, crystals with measured deficiencies of Sr and Mn were reported to exhibit a FM moment that scales with the deficiency level \cite{Liu_NatMat2017}. However, neutron scattering on samples from the same group concluded that the FM order occurs as a minority phase, while the majority phase is antiferromagnetic with no evidence of spin canting \cite{Zhang_PRB2019}; later inelastic neutron scattering similarly found that the spin dynamics are well described by a purely AFM model, with no ferromagnetic component \cite{Cai2020,Ning_PRB2024}. A separate group demonstrated that SrMnSb$_2$ crystals can be grown with no measurable FM moment \cite{Liu_PRB2019}; furthermore, a powder diffraction study showed that heating above $\sim 500$~K leads to the appearance of small amounts of SrO and Sb, as well as a ferromagnetic response, suggesting a link to sample decomposition.

A similar controversy has emerged in \ymb\ \cite{Wang_2016}. Angle-resolved photoemission spectroscopy combined with magneto-optical microscopy reported evidence for TRS breaking at the crystal surface, interpreted in terms of an $\sim 10^\circ$ canting of Mn moments \cite{Borisenko_2019}. TRS breaking was also inferred from a large anomalous Nernst signal in a crystal with a weak FM moment; however, the implied canting angle was only $\sim 0.02^\circ$ \cite{Pan_NatMat2022}, effectively indistinguishable from zero within realistic material constraints. Neutron diffraction measurements \cite{Soh_2019,Zaliznyak_2017} found no evidence for the spin canting proposed in Ref.~\onlinecite{Borisenko_2019}, calling into question a bulk Weyl semimetal state in this material. Optical-conductivity studies have likewise disagreed on whether Weyl fermions are essential \cite{Chinotti_2016} or not \cite{Chaudhuri_2017} for interpreting the data.

{Taken together, these results highlight that bulk TRS breaking via magnetic moment canting is difficult to establish unambiguously in the 112 family. This motivates direct diffraction tests~\cite{Soh_2019,Soh_PRB2021}, but also raises the possibility that some reported low-temperature anomalies instead originate from distinct intrinsic symmetry-lowering instabilities of the square-net Dirac layers, as we find here.}

{In \cmb, the transition at $T^{*}=46(2)$~K corresponds to} a coupled structural and magnetic symmetry lowering consistent with a Peierls-type BOW/CDW distortion in the Dirac-electron Bi layer, as originally predicted by Hoffmann and co-workers \cite{TremelHoffman_JACS1987,PapoianHoffman_AngewChem2000}. This transition is manifested by the appearance below $T^{*}$ of $(H,0,L+\tfrac{1}{2})$ superlattice Bragg reflections with integer $H\neq 0$ and $L$ (indexed with respect to the parent $P4/nmm$ cell) in both neutron and x-ray diffraction, together with orthorhombic splitting of the fundamental $(H,0,L)/(0,H,L)$ peaks in x-ray diffraction; the twinned reflections separate in the orthorhombic phase. The superlattice peaks contain comparable nuclear and magnetic contributions, consistent with anti-phase shearing displacements of the magnetic layers. 

Overall, CaMnBi$_2$ exhibits three distinct phases in terms of magnetic and crystal structure: a high-temperature tetragonal paramagnetic phase ($T > T_\mathrm{N}$), a tetragonal C-type antiferromagnetic phase ($T^{*} < T < T_\mathrm{N}$), and a low-temperature orthorhombic Peierls/BOW phase with little impact on C-type antiferromagnetism ($T < T^{*}$) [Fig.~\ref{CMS}]. Although the orthorhombic symmetry in principle allows spin canting and the associated TRS breaking that could support a Weyl state at $T < T^{*}$, any such weak canting, if present, is beyond our experimental detection limit.

Rietveld refinements of the polarized and unpolarized neutron diffraction data show that below $T^{*}$, \cmb\ transforms from the high-temperature tetragonal $P4/nmm$ lattice with magnetic space group $P4^\prime /n^\prime m^\prime m$ to an orthorhombic $Pcmn$ lattice with magnetic space group $Pc^\prime m^\prime n^\prime$. Between the two candidate irreducible representations, $Z_5^-$ and $Z_5^+$, of the low-temperature orthorhombic structure compatible with the symmetry constraints of the observed $P4/nmm \rightarrow Pcmn$ transition, the $Z_5^-$ mode, corresponding to a zigzag BOW in the Bi1 Dirac layer, is consistently, though modestly, favored by the neutron refinements. DFT likewise finds a slightly lower energy for the $Z_5^-$ configuration; however, the displacement amplitudes are near the limit of theoretical accuracy, and the undistorted tetragonal $P4/nmm$ structure remains the global minimum. The $Z_5^-$ IR, corresponding to a Peierls-type BOW/CDW distortion in the Dirac-electron Bi1 layer [Fig.~\ref{Fig:IPD}(a),(b)], is further supported by several experimental observations.

Although the superlattice reflections observed by polarized neutron diffraction contain comparable structural and magnetic components, they develop continuously through a second-order phase transition, as evidenced by the similar order-parameter-like temperature dependence of both the superlattice peak intensities and the lattice distortion parameter [Figs.~\ref{SPUP}, \ref{xray}, and \ref{SPP}]. This behavior indicates a single primary order parameter. {While we do not have enough data within a narrow critical range of $\sim 10\%$ of $T_c$ to reliably evaluate the critical behavior, making the extracted critical exponents somewhat phenomenological, analyses using variable fitting ranges (Appendix~\ref{App:CriticalExponent}) consistently yield small exponents, $\beta\sim0.2$, indicative of a quasi-two-dimensional character of the observed Peierls transition.} The small critical exponent is compatible with 2D Ising-like critical behavior and supports an electronically driven 2D Peierls/BOW transition in a layered system, which can be mapped onto an effective quasi-2D Ising model \cite{ZacharZaliznyak_PRL2003,Zaliznyak_PRB2012}. This contrasts with the 3D critical behavior of the antiferromagnetic order parameter at $T_\mathrm{N}$ shown in Fig.~\ref{Fig:Neel_OP}. An electronic Peierls mechanism illustrated in Fig.~\ref{CMS}(d),(e) is also consistent with the partial Fermi-surface gapping inferred from optical-reflectivity measurements \cite{Corasaniti_2019,Yang_2020} and reproduced in our DFT calculations [Fig.~\ref{Fig:DFT_Fig2S_bands}].

In the $Z_5^+$ model, which involves a BOW in the Mn–Bi layer, the magnetic contribution to the superlattice peaks arises solely from spin canting. This scenario would involve both lattice and magnetic symmetry breaking and hence two coupled order parameters, implying a first-order phase transition. Indeed, such a BOW has been observed in FeTe, where orbital dimerization in the Fe-Te layer occurs through a first-order phase transition and produces pronounced changes in Fe magnetism, manifested by a substantial change in Fe magnetic moment and magnetic susceptibility \cite{Zaliznyak_PRL2011,Zaliznyak_PRB2012,Fobes_PRL2014}. In contrast, Mn magnetism in \cmb\ appears largely insensitive to the transition at $T^{*}$. The refined magnetic moment at 50~K, $\mu_{\mathrm{Mn}}(z) = 3.68(5) \mu_B$, is only slightly smaller than at 5~K, $\mu_{\mathrm{Mn}}(z) = 3.97(2) \mu_B$, consistent with the expected temperature evolution of the magnetic order parameter, and the magnetic susceptibility shows no discernible anomaly at $T^{*}$ [Fig.~\ref{MR}(a)].

Finally, a similar structural distortion with a zigzag BOW corresponding to the $Z_5^-$ IR of the orthorhombic group, and consistent with the predicted Peierls-type BOW/CDW instability in the Dirac-electron layer \cite{TremelHoffman_JACS1987,PapoianHoffman_AngewChem2000}, has been observed in several related Sr/Ca/EuMnSb$_2$ materials \cite{You_CurrApplPhys2019,Zhang_PRB2019,Ning_PRB2024,Brechtel_JLessCommMet1981,He_PRB2017,Qiu_PRB2018,Rong_PRB2021,Sakai_PRB2020,Liu_NatComm2021,Yoshizawa_PRB2022,Wilde_PRB2022}. Most recently, it was reported in the sister compound YbMnSb$_2$ \cite{Bhoi_PRB2025}, where the orthorhombic distortion, although much smaller than in other antimonides, is about 50\% larger than in \cmb. According to DFT calculations, this distortion leads to a reconstruction of the electronic structure similar to what we find in \cmb\ for atomic displacements approximately twice as large as those observed experimentally. Both results are consistent with an electronic Peierls mechanism for the distortion.

In YbMnSb$_2$, the distortion persists to high temperatures, well above magnetic ordering at $T_\mathrm{N} \approx 350$~K, similar to other antimonides \cite{Bhoi_PRB2025}. This led the authors to stipulate that YbMnSb$_2$ crystallizes in an orthorhombic structure whose reduced symmetry governs its electronic properties at all temperatures. In contrast, we find that in \cmb, both crystal and magnetic symmetry are lowered only at low temperature, within the magnetically ordered phase, placing the transition squarely in the regime of low-energy physics governed by electronic interactions. Furthermore, our preliminary measurements \cite{Hu_PRB2023,Hu_unpublished2025} indicate that the distortion in YbMnSb$_2$ disappears upon heating to higher temperatures, well below melting, suggesting that it too originates from the same Peierls-type electronic instability of Dirac electrons.

\section{Summary and Conclusions}
\label{Sec:Summary}

{Our combined polarized and unpolarized neutron diffraction and x-ray measurements on single-crystal CaMnBi$_2$, supported by DFT, identify the structural and magnetic changes underlying the resistive anomaly at $T^{}\approx 50$~K \cite{Wang_2012,He_2012,Corasaniti_2019,Yang_2020}. We show that $T^{}=46(2)$~K corresponds to a coupled structural and magnetic symmetry lowering consistent with an electronically driven Peierls-type charge/bond-ordering instability of the Dirac-electron Bi square net \cite{TremelHoffman_JACS1987,PapoianHoffman_AngewChem2000}.}

{Our neutron measurements rule out the previously proposed large ($\approx 10^\circ$) \emph{uniform} canting of Mn moments in \cmb; any uniform canting, if present, is $\lesssim 2^\circ$. Therefore, the resistivity/optical anomalies at $T^{*}$ do not originate from spin canting or weak ferromagnetism and the associated Weyl semimetallic state \cite{Wang_2011,He_2012,Corasaniti_2019,Yang_2020}.}

Taken together, our findings establish that the transition at $T^{*}$ in CaMnBi$_2$ represents a rare realization of a two-dimensional Peierls instability toward bond ordering in a Dirac-electron square-net system \cite{TremelHoffman_JACS1987,PapoianHoffman_AngewChem2000}. The subtle structural distortion, with negligible impact on Mn antiferromagnetism, underscores the primarily Dirac-electron origin of the transition and highlights the broader role of Peierls-type lattice instabilities in shaping the ground states of topological semimetals. Although such transitions have long been studied theoretically \cite{TremelHoffman_JACS1987,PapoianHoffman_AngewChem2000,Ono_JPSJ2000,Aryal_PRB2022,Alekseev_PRB2026}, experimental realizations have remained scarce or lacking.

Our results not only bridge this gap, but also carry important implications for materials design. The observed electronic instability of Dirac electrons is sensitive to the Fermi level, which can be tuned compositionally and likely explains the absence of the transition in samples from Ref.~\onlinecite{Guo_PRB2014}. The transition temperature $T^{*}$ associated with the resistivity anomaly increases by nearly a factor of two in Ca$_{1-x}$Na$_{x}$MnBi$_{2}$ with $x=0.05$ \cite{Corasaniti_2019,Yang_2020}. Likewise, systematic variation of the Bi/Sb content in (Ca,Yb)MnBi$_{1-x}$Sb$_x$ compounds, where Bi-based members show no distortion or only a weak low-temperature Peierls distortion as in \cmb, while antimonides exhibit a well-developed distortion, offers a promising route to further experimental and theoretical exploration of electronic instabilities and realizing tunable functionalities in Dirac semimetals.

\section{Acknowledgement}

We thank Tyler J. Slade and Efrain Rodriguez for drawing our attention to the work of Hoffmann on Sb and Bi square-net compounds. The work at Brookhaven was supported by the Office of Basic Energy Sciences, U.S. Department of Energy (DOE) under Contract No. DE-SC0012704. This research used resources at the High Flux Isotope Reactor, a DOE Office of Science User Facility operated by the Oak Ridge National Laboratory. The beam time was allocated to PTAX proposal IPTS-27767 and DEMAND proposal IPTS-32665. Work at Ames National Laboratory was supported by the U.\,S.\ Department of Energy (DOE), Basic Energy Sciences, Division of Materials Sciences \& Engineering, under Contract No.\ DE-AC$02$-$07$CH$11358$.

\appendix

\section{Experimental limit on the spin-canting contribution to (0, 0, L) peaks in Ca$_{1-x}$Na$_x$MnBi$_2$}
\label{App:SCCNMB}
\begin{figure}[h!]
\centering
\includegraphics[scale=0.8]{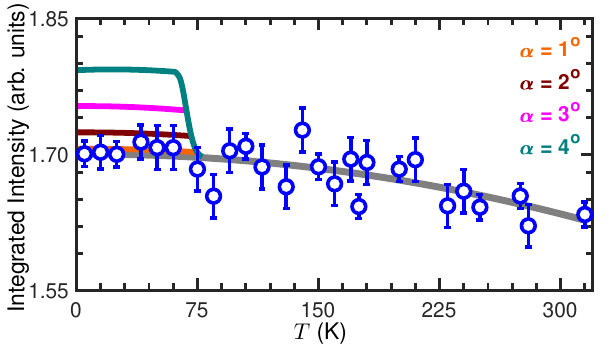}
\caption{Temperature dependence of the integrated intensity of $(0, 0, 2)$ peak in \cnmb\ ($x = 0.05$) crystal. Lines show a polynomial fit of the $(0, 0, 2)$ integrated intensity (gray) and the calculated intensity enhancement due to spin-canting for the tilt angles $\alpha = 1{\degree}$ (orange), $2{\degree}$ (brown),  $3{\degree}$ (magenta), and $4{\degree}$ (teal).}
\label{00L_FM_0p05}
\end{figure}

Figure~\ref{00L_FM_0p05} shows the temperature dependence of the integrated intensity of the $(0, 0, 2)$ peak measured for \cnmb\ $x = 0.05$ crystal on SPINS at NCNR. The lines correspond to the polynomial fit of $(0, 0, 2)$ intensity (gray) and calculated intensities for canting angle $\alpha = 1{\degree}-4{\degree}$ using the methodology discussed in Ref.~\citenum{Soh_2019}. From the plot and our calculation it is clear that the intensity enhancement should be readily detectable for the canting angle $> 2{\degree}$, thereby implying that canting in \cnmb\ $x = 0.05$ should be $\leq 2{\degree}$, as opposed to $10{\degree}$ canting stipulated in Ref.~\citenum{Yang_2020}.

\section{Superlattice peaks in different Brillouin zones}
\label{App:SLPDBZ}
\begin{figure}[h!]
\centering
\includegraphics[scale=0.73]{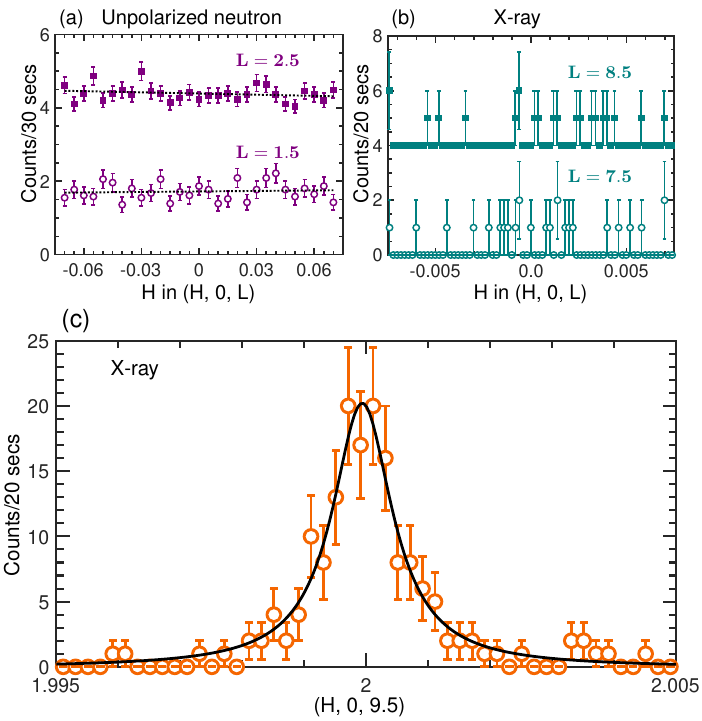}
\caption{(a) and (b) $H$-scans through $(0, 0, L + 0.5)$ superlattice peaks measured with unpolarized neutrons at SPINS and four-circle X-ray diffractometer at Ames National Laboratory, respectively. (c) $H$-scan through (2, 0, 9.5) superlattice peak measured using X-ray diffraction. Solid black line is a fit with the Lorentzian lineshape.}
\label{SS_H0andH2}
\end{figure}

In both unpolarized neutron and X-ray diffraction measurements superlattice peaks at reciprocal ($P4/nmm$) lattice positions $(H, 0, L + 0.5)$, $L$ is an integer, in different Brillouin zones were measured (Fig.~\ref{SS_H0andH2}). X-ray diffraction found the superlattice peaks present at $(2,\ 0,\ \dfrac{2n+1}{2})$ positions, $n$ is an integer, as shown in Fig.~\ref{SS_H0andH2}(c); other peaks measured were (2, 0, 7.5) and (2, 0, 8.5).  Superlattice peaks with $H = 0$ were absent in both neutron and X-ray measurements, as shown in Fig.~\ref{SS_H0andH2}, (a) and (b). Zero intensity of $(0, 0, L + 0.5)$ superlattice peaks indicates that atomic displacements involved in the symmetry-lowering structural transition are perpendicular to the $c$-axis (i.e, are in the $ab$-planes), such as the layer-shearing distortions found in YbMnSb$_2$ \cite{Bhoi_PRB2025,Hu_PRB2023,Hu_unpublished2025}.

\section{Critical-exponent analysis}
\label{App:CriticalExponent}

\begin{figure}[h!]
\centering
\includegraphics[width=1.0\columnwidth]{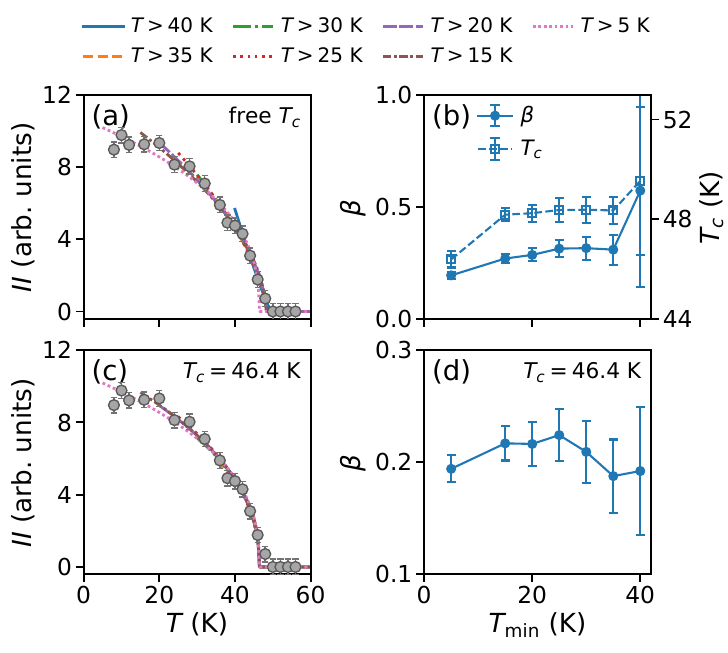}
\caption{Temperature dependence of the (1, 0, 0.5) superlattice reflection measured by neutron diffraction shown in Fig.~\ref{SPUP}(e) (symbols), with order-parameter fits over different temperature windows with lower temperature cutoffs $T_{\min}=5$, 15, 20, 25, 30, 35, and 40~K (lines), varying both $T_c=T^*$ and $\beta$ in (a) and only $\beta$ with fixed $T_c=46.4$~K in (c). The resulting dependencies of the fit parameters on the fitting window are shown in (b) and (d), respectively.}
\label{Fig5e_critical_analysis}
\end{figure}

\begin{figure}[h!]
\centering
\includegraphics[width=1.0\columnwidth]{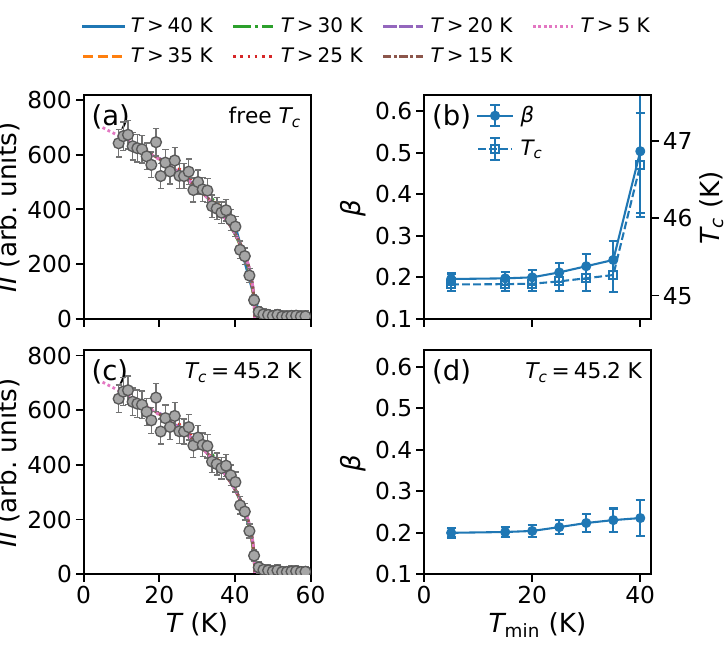}
\caption{Temperature dependence of the $(1,0,8.5)$ superlattice reflection measured by x-ray diffraction shown in Fig.~\ref{xray}(b) (symbols), with order-parameter fits over different temperature windows with lower temperature cutoffs $T_{\min}=5$, 15, 20, 25, 30, 35, and 40~K (lines), varying both $T_c=T^*$ and $\beta$ in (a) and only $\beta$ with fixed $T_c=45.2$~K in (c). The resulting dependencies of the fit parameters on the fitting window are shown in (b) and (d), respectively. }
\label{Fig6b_critical_analysis}
\end{figure}

\begin{figure}[h!]
\centering
\includegraphics[width=1.0\columnwidth]{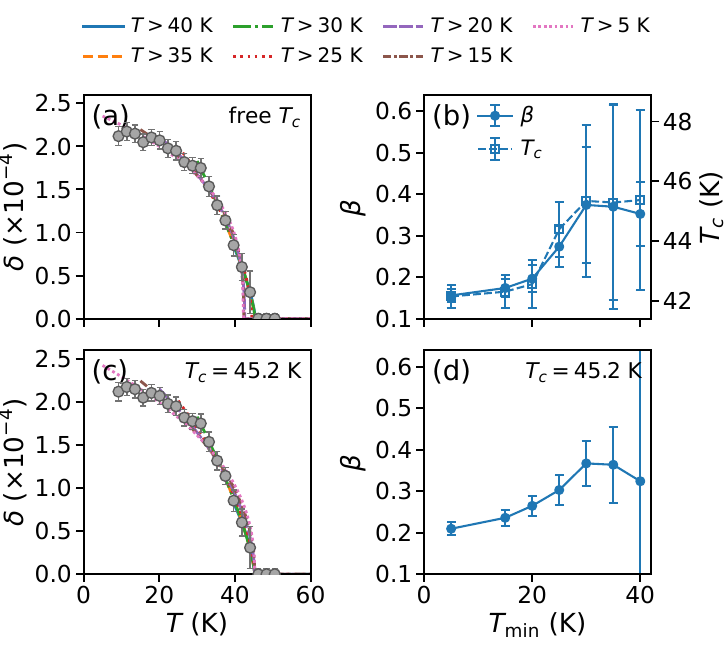}
\caption{Temperature dependence of the orthorhombic splitting $\delta$ of the $(3,0,8)$ lattice reflection measured by x-ray diffraction shown in Fig.~\ref{xray}(d) (symbols), with order-parameter fits over different temperature windows with lower temperature cutoffs $T_{\min}=5$, 15, 20, 25, 30, 35, and 40~K (lines), varying both $T_c=T^*$ and $\beta$ in (a) and only $\beta$ with fixed $T_c=45.2$~K in (c). The resulting dependencies of the fit parameters on the fitting window are shown in (b) and (d), respectively.}
\label{Fig6d_critical_analysis}
\end{figure}

\begin{table}[h!]
\caption{\label{Table:critical_exponent_analysis}
Summary of the critical-law fits for different temperature windows. The values are the unweighted mean and sample standard deviation obtained from the six fitting ranges with lower temperature cutoffs $T_{\min}=5$, 15, 20, 25, 30, and 35~K; fits in the narrowest range $T>40$~K contain too few data points to reliably constrain the fit parameters and are excluded. For the fixed-$T_c$ analyses, only $\beta$ is averaged.}

\begin{ruledtabular}
\begin{tabular}{lcc}
Data set / analysis & $T_c$ (K) & $\beta$ \\
\hline
Neutron $II$, Fig.~5(e), $T_c$ free & $48.0(8)$ & $0.28(5)$ \\
Neutron $II$, Fig.~5(e), $T_c=46.4$~K & --- & $0.208(14)$ \\
X-ray $II$, Fig.~6(b), $T_c$ free & $45.19(5)$ & $0.21(2)$ \\
X-ray $II$, Fig.~6(b), $T_c=45.2$~K & --- & $0.21(1)$ \\
X-ray $\delta$, Fig.~6(d), $T_c$ free & $43.67(1.51)$ & $0.26(1)$ \\
X-ray $\delta$, Fig.~6(d), $T_c=45.2$~K & --- & $0.29(7)$ \\
\end{tabular}
\end{ruledtabular}
\end{table}

While we do not have enough data points within a narrow critical range of $\sim 10\%$ of $T_c$ to obtain reliable order-parameter fits within this range, making the extracted critical exponents somewhat phenomenological, we assessed their sensitivity to the temperature range by repeating the fits using progressively restricted fitting windows, with lower cutoffs $T_{\min}=5$, 15, 20, 25, 30, 35, and 40~K. The integrated neutron and x-ray superlattice intensities, $II$, and the orthorhombic splitting, $\delta$, were fitted to $A(1-T/T_c)^{2\beta}$, assuming that these quantities are proportional to the square of the primary order parameter. The results are summarized in Figs.~\ref{Fig5e_critical_analysis}, \ref{Fig6b_critical_analysis}, and \ref{Fig6d_critical_analysis}. The narrowest $T>40$~K fits contain too few points to constrain all fitting parameters reliably and were therefore excluded when calculating the averages in Table~\ref{Table:critical_exponent_analysis}. The quoted uncertainties in this table are sample standard deviations among the six fitting windows and thus characterize the sensitivity to the choice of fitting range rather than the statistical uncertainty of an individual fit.

The x-ray superlattice intensity in Fig.~\ref{xray}(b) gives the most stable result, with $T_c=45.186(49)$~K and $\beta=0.212(19)$ for free $T_c$ and $\beta=0.212(13)$ for fixed $T_c=45.2$~K, averaged over the fitting windows. The neutron intensity in Fig.~\ref{SPUP}(e) shows somewhat stronger covariance between $T_c$ and $\beta$; fixing $T_c=46.4$~K gives a comparatively stable $\beta=0.208(14)$. The orthorhombic splitting in Fig.~\ref{xray}(d) is considerably more sensitive to the fitting range, yielding $\beta=0.257(97)$ when $T_c$ is free and $\beta=0.291(66)$ for $T_c=45.2$~K fixed by the x-ray superlattice intensity. In all analyses, the extracted exponents are consistently smaller than those expected for three-dimensional models, which we take as indicative of a quasi-two-dimensional character of the observed Peierls transition.

\section{Symmetry reduction in the low-temperature L = 0.5 superlattice}
\label{App:ISODISTORT}

\begin{table}[h!]
	\caption{Isotropy subgroups and irreducible representations (IRs) for the $P4/nmm$ space group of \cmb\ with the \textbf{k}-vector of (0, 0, 0.5). “Yes” and “No” indicates whether or not a phase transition from the parent to the subgroup symmetry is allowed to be continuous (second order) \cite{Stokes_2016}. \label{IRSG}}
 	\begin{ruledtabular}
		\begin{tabular}{|c|c|c|}
            IR & Isotropy Subgroups & Continuous Transition\\
              &  & Allowed?\\
            \hline
 			$Z^{+}_1$ & $P4/nmm$ & Yes\\
            \hline
 		    $Z^{+}_2$ & $P4_{2}/nmc$ & Yes\\
            \hline
                        & $Cmca$ & Yes\\
             $Z^{+}_5$  & $Pnma$ & Yes\\
                         & $P2_{1}c$ & No\\
            \hline
            $Z^{-}_3$ & $P4/nmm$ & Yes\\
            \hline
            $Z^{-}_4$ & $P4_{2}/nmc$ & Yes\\
            \hline
                        & $Cmca$ & Yes\\
             $Z^{-}_5$  & $Pnma$ & Yes\\
                         & $P2_{1}c$ & No\\
 		\end{tabular}
 	\end{ruledtabular}
  \end{table}

The observed superlattice distortion transition at $T^{*}$ is of a second order, so a group-subgroup relationship can be exploited to infer the low-temperature crystal symmetry at $T < T^{*}_\mathrm{s}$. Therefore, we used ISODISTORT\cite{Campbell_2006} and ISOSUBGROUP\cite{Stokes_2016} tools of the ISOTROPY software suite \cite{Stokes_iso}where the $P4/nmm$ space group of \cmb\ with Mn at $2a$ Wyckoff site and \textbf{k}-vector of $(0, 0, 0.5)$ are used; $(0, 0, 0.5)$ and $(1, 0, 0.5)$ are equivalent. In this case, we obtain six irreducible representations (IRs, or IRs) for the special Z-point, $Z^{+}_1$, $Z^{+}_2$, $Z^{+}_5$, $Z^{-}_3$, $Z^{-}_4$, and $Z^{-}_5$.

\begin{table}[]
	\caption{Extinction rules for the superlatttice peaks corresponding to the Isotropy subgroups $Cmca$ and $Pnma$ of IRs $Z_5^+$ and $Z_5^-$ for the $P4/nmm$ space group of \cmb\ with the \textbf{k}-vector of (0, 0, 0.5). ``Yes'' and ``No'' indicate the presence and the absence of the X-ray diffraction peaks, considering one or any combination of the allowed atomic displacement modes for Ca, Mn, Bi1 and Bi2. ``Odd'' and ``Even'' mean odd and even integer values for $H$ and $K$, $L = \dfrac{2n+1}{2}$ is half-integer.}
\label{SGCP}
 	\begin{ruledtabular}
		\begin{tabular}{|l|l|c|c|c|c|c|}
            \multirow{2}{*}{IRs} &
            \multirow{2}{*}{Subgroups} &
            \multicolumn{2}{c|}{$(H, 0, L)$} &
            \multicolumn{2}{c|}{$(0, K, L)$} &
            $(0 0 L)$\\
              & & {Odd} & {Even} & {Odd} & {Even} & \\
            \hline
             $Z^{+}_5$  & $Cmca$ & Yes & Yes & Yes & Yes & No\\
             \hline
             $Z^{+}_5$  & $Pnma$ & Yes & Yes & No & No & No\\
            \hline
            $Z^{-}_5$   & $Cmca$ & Yes & Yes & Yes & Yes & No \\
             \hline
             $Z^{-}_5$  & $Pnma$ & Yes & Yes & No & No & No\\
            \hline
             &Experiment & Yes & Yes & No & No & No\\
 		\end{tabular}
 	\end{ruledtabular}
  \end{table}

Table ~\ref{IRSG} lists the ISOTROPY subgroups for all 6 IRs along with the information about whether the transition from parent to the subgroup is continuous or not. Out of the six IRs, $Z^{+}_1$, $Z^{+}_2$, $Z^{-}_3$, and $Z^{-}_4$ preserve the tetragonal crystal structure, hence are inconsistent with the observed splitting. Among the remaining subgroups, the monoclinic subgroup $P2_{1}c$ and the orthorhombic subgroup $Cmca$ corresponding to IRs $Z^{+}_5$ and $Z^{-}_5$ can be disregarded. $P2_{1}c$ is not a maximal subgroup and is inconsistent with the observed continuous or second order phase transition, as listed in the Table~\ref{IRSG}. On the other hand, for the orthorhombic subgroup $Cmca$, where $a_\mathrm{orth}$ and $b_\mathrm{orth}$ are rotated by $45^\degree$ with respect to tetragonal $a$ and $b$ of $P4/nmm$, the splitting of the peaks would be expected along $[1, 1, 0]$ $P4/nmm$ direction, in disagreement with our experimental observation of splitting along the $[1, 0, 0]$ direction.

In addition, the ISODISTORT interactive visualization tools were utilized to simulate and examine the single crystal X-ray diffraction patterns across various reciprocal planes and for different modes of atomic displacements. The results obtained from these simulations are tabulated in Table~\ref{SGCP}. The Table reveals that only for the orthorhombic subgroup $Pnma$ there are displacement modes that produce superlattice peaks consistent with the experimental observations. On the other hand, no such displacement mode or a combination of the displacement modes exist for $Cmca$. Moreover, transition from $P4/nmm$ to $Pcmn$ ($Pnma$) allows the splitting of the Bragg peaks along [1, 0, 0] due to formation of twin domains. The difference in $Pcmn$ ($Pnma$) subgroups corresponding to two IRs $Z^{-}_5$ and $Z^{+}_5$ is the shift in the origin, by (0, 0, 0) and (0, 0, 1/2), respectively, with respect to high-temperature $P4/nmm$ parent space group.

In summary, our findings reveal that \cmb\ undergoes a second order structural transition from tetragonal $P4/nmm$ to orthorhombic $Pcmn$ ($Pnma$) below $T^{*}_\mathrm{s} = 46 (1)$~K, corresponding to the resistive anomaly. Nonetheless, the results do not definitively dismiss the potential magnetic origin or magnetic nature of the observed superlattice peaks.

\section{Polarized-Neutron TAS Refinement of the Low-$T$ Crystal and Magnetic Structure}
\label{App:TAS_refinement}
Two space groups were considered for the crystal structure refinement at 5~K: tetragonal $P4/nmm$ and orthorhombic $Pcmn$ (standard setting: $Pnma$), as discussed in the main text. Because only a limited number of Bragg peaks were measured, we used isotropic atomic displacement parameters (ADPs) $B_\mathrm{iso}$ in the refinements. To further reduce the number of free parameters, and consistent with the observed absence of $H = 0$ superlattice reflection, we constrained the atomic displacements to the $ab$-plane by fixing the $z$-coordinates of the Mn and Bi1 atoms to their nominal positions.

\begin{figure}[h!]
\centering
\includegraphics[width=0.9\columnwidth]{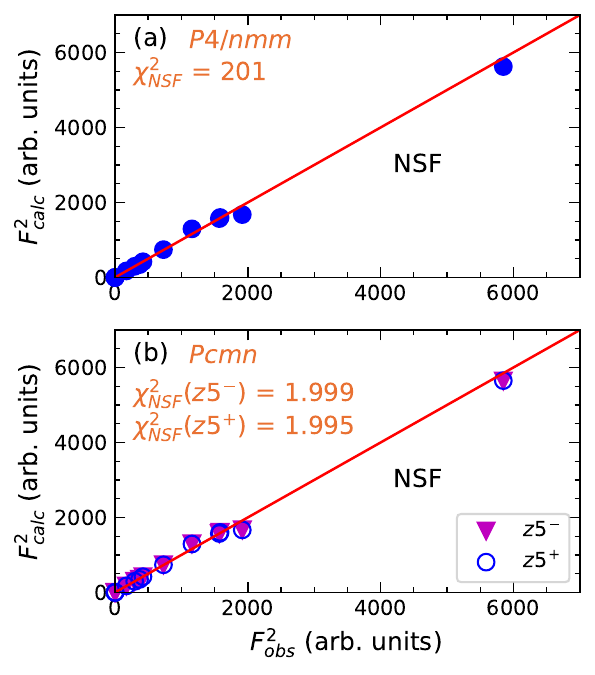}
\caption{Structural Rietveld refinement of \cmb\ using lattice nuclear Bragg peak intensities measured at 5~K in NSF channel for (a) $P4/nmm$ and (b) $Pcmn$ (standard setting: $Pnma$) crystal symmetry. Plots compare calculated and observed intensities, expressed as the squared structure factor (arbitrary units). Magenta triangles and blue open circles in (b) correspond to results for $Z_5^{-}$ and $Z_5^{+}$, respectively.}
\label{Fig:TAS_NSF_refinement}
\end{figure}
%

\begin{table}[h!b!]
\caption{\textbf{TAS NSF-only refinements.} Crystal structures for the $Z^{-}_5$ and $Z^{+}_5$ IRs in the orthorhombic $Pcmn$ setting, obtained from Rietveld refinement of integrated Bragg intensities measured in the NSF channel at 5~K. Red values denote unphysical or poorly constrained parameters. Lattice parameters: $a = 4.483$~\AA, $b = 4.481$~\AA, $c = 21.968$~\AA. \label{TAS_NSFonly_refinement}}
\begin{ruledtabular}
\begin{tabular}{ccccc}
\multicolumn{5}{c}{$Z^{-}_5$ ; $Pcmn$; $\chi^2 = 1.999$} \\
\hline
Atom & $x$ & $y$ & $z$ & $B_\mathrm{iso}$ (\AA$^2$)\\
\hline
Ca  & 0.7702(4) & 0.2500 & 0.6174(4) & 1.089(200)\\
Mn  & 0.2716(89) & 0.2500 & 0.7500 & 1.018(379)\\
Bi1 & \textcolor{red}{0.2466(2)} & 0.2500 & 0.5000 & 1.352(228)\\
Bi2 & 0.7590(2) & 0.2500 & 0.8306(2) & 1.820(332)\\
\end{tabular}
\end{ruledtabular}
\begin{ruledtabular}
\begin{tabular}{ccccc}
\multicolumn{5}{c}{$Z^{+}_5$ ; $Pcmn$; $\chi^2 = 1.995$} \\
\hline
Atom & $x$ & $y$ & $z$ & $B_\mathrm{iso}$ (\AA$^2$)\\
\hline
Ca  & 0.7648(4) & 0.2500 & 0.8676(4) & 1.215(285)\\
Mn  & \textcolor{red}{0.2542(4)} & 0.2500 & 0.0000 & 1.311(407)\\
Bi1 & 0.2493(998) & 0.2500 & 0.2500 & 1.348(176)\\
Bi2 & 0.7332(2) & 0.2500 & 0.0806(2) & 1.658(199)\\
\end{tabular}
\end{ruledtabular}
\end{table}

We first refined the crystal lattice using nuclear Bragg peak intensities measured by polarized TAS diffraction in the NSF channel, including the superlattice peaks. Figure~\ref{Fig:TAS_NSF_refinement} summarizes the NSF-only lattice refinements. The calculated and measured nuclear intensities agree well for the orthorhombic $Pcmn$ model, yielding $\chi^2 \approx 2$, whereas the tetragonal $P4/nmm$ model gives $\chi^2 \approx 200$, indicating a significant discrepancy. The large $\chi^2$ for $P4/nmm$ arises primarily from the observed half-integer ($L = n + \tfrac{1}{2}$) superlattice peaks; when these reflections are excluded, the fit improves to $\chi^2 \approx 2.6$. These results demonstrate that the low-temperature crystal structure is orthorhombic $Pcmn$, consistent with the x-ray diffraction results discussed in Section~\ref{XRD}.

The refined structural parameters for the two IRs ($Z^{-}_5$ and $Z^{+}_5$) are listed in Table~\ref{TAS_NSFonly_refinement}. Both models provide equally good fits to the NSF data. The refinements reveal several nearly equal-depth $\chi^2$ minima in the range $x \in [0.2466, 0.258]$ for the displacement of Bi1 along the $a$-axis in $Z^{-}_5$, whereas for Mn displacement in $Z^{+}_5$, an additional broad $\chi^2$ minimum is found near $x = 0.27$. This indicates that the refined values are poorly constrained due to the very limited number of Bragg reflections accessible in our polarized TAS diffraction measurements, which also results in overestimated isotropic ADPs ($B_\mathrm{iso}$). Such poorly constrained parameters are shown in red in Table~\ref{TAS_NSFonly_refinement}.


\begin{figure}[h!t!]
\centering
\includegraphics[width=0.9\columnwidth]{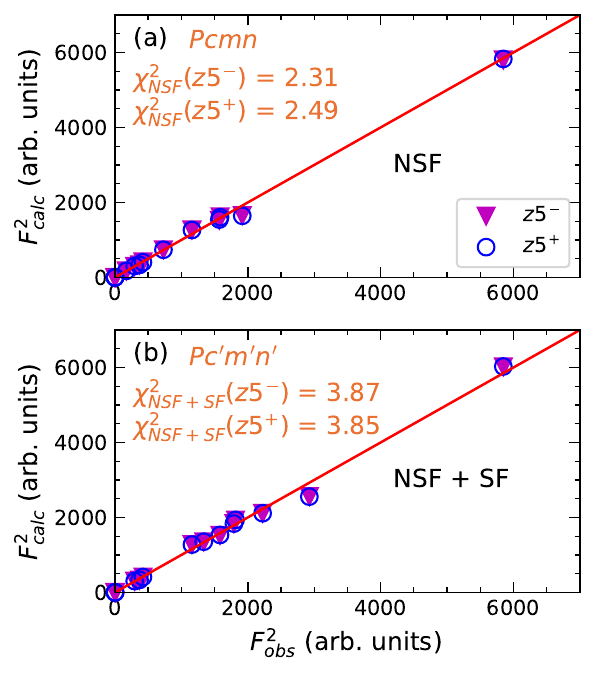}
\caption{Structural (a) and magnetic (b) refinements of \cmb\ using NSF and combined NSF and SF Bragg intensities measured at 5~K. Calculated and observed intensities are compared as squared structure factors (arbitrary units). Structural refinement (a) was performed with crystal symmetry $Pcmn$, and magnetic refinement (b) with magnetic symmetry $Pc'm'n'$ (standard setting: $Pn'm'a'$). Magenta triangles correspond to results for $Z^{-}_5$, and blue open circles correspond to $Z^{+}_5$.}
\label{TAS_individual_fixBiso}
\end{figure}

Next, we attempted to refine the magnetic structure using the combined NSF+SF Bragg intensities. We found that the magnetic structure could not be reliably refined when the atomic positions were fixed to the NSF-refined nuclear lattice structures listed in Table~\ref{TAS_NSFonly_refinement}. Consequently, we refined both the magnetic and lattice structures for each $Pcmn$ IR ($Z^{-}_5$ and $Z^{+}_5$) using the NSF+SF data. To avoid unphysically large ADPs, $B_\mathrm{iso}$ values were fixed to physically reasonable values, ultimately adopting those obtained from subsequent unpolarized neutron refinement at the HB-3A DEMAND diffractometer (App.~\ref{App:HB3A_refinement}), supplemented with the rescaled TAS superlattice peaks. The NSF-only refinement was also repeated using these fixed $B_\mathrm{iso}$ values.  Figure~\ref{TAS_individual_fixBiso} presents a summary of these refinements, and the corresponding structural parameters are listed in Table~\ref{TAS_individual_refinement}.

\begin{table}[h!]
\caption{\textbf{NSF and NSF+SF TAS refinements with fixed ADPs.} $B_\mathrm{iso}$ were fixed to the values obtained from HB-3A refinements (App.~\ref{App:HB3A_refinement}), supplemented with the 3 rescaled TAS superlattice peaks. Although NSF+SF intensities together are effectively equivalent to unpolarized data, fitting the lattice and magnetic structure simultaneously imposes additional constraints. Lattice parameters: $a = 4.483$~\AA, $b = 4.481$~\AA, $c = 21.968$~\AA. \label{TAS_individual_refinement}}
\begin{ruledtabular}
\begin{tabular}{ccccc}
$Z^{-}_5$ & \multicolumn{3}{c}{ NSF-only } & $\chi^2 = 2.31$\\
\hline
Atom & $x$ & $y$ & $z$ &$B_\mathrm{iso}$ (\AA$^2$)\\
\hline
Ca  & 0.7707(5) & 0.2500 & 0.6182(5) & 0.593\\
Mn  & 0.2719(91) & 0.2500 & 0.7500 & 0.713\\
Bi1 & \textcolor{red}{0.2465(2)} & 0.2500 & 0.5000 & 0.637\\
Bi2 & 0.7591(2) & 0.2500 & 0.8306(2) & 0.609\\
\end{tabular}
\end{ruledtabular}
\begin{ruledtabular}
\begin{tabular}{ccccc}
& \multicolumn{3}{c}{ NSF + SF } & $\chi^2 = 3.87$\\
\hline
Atom & $x$ & $y$ & $z$ &$B_\mathrm{iso}$ (\AA$^2$)\\
\hline
Ca  & 0.7722(7) & 0.2500 & 0.6188(7) & 0.593\\
Mn  & 0.2504(51) & 0.2500 & 0.7500 & 0.713\\
Bi1 & \textcolor{red}{0.2458(3)} & 0.2500 & 0.5000 & 0.637\\
Bi2 & 0.7599(3) & 0.2500 & 0.8305(2) & 0.609\\
\hline
\multicolumn{5}{c}{$\mu_{Mn}(x) = N/A$, $\mu_{Mn}(z) = 4.23(7)\mu_B$} \\
\end{tabular}
\end{ruledtabular}
\begin{ruledtabular}
\begin{tabular}{ccccc}
$Z^{+}_5$& \multicolumn{3}{c}{ NSF-only } & $\chi^2 = 2.49$\\
\hline
Atom & $x$ & $y$ & $z$ &$B_\mathrm{iso}$ (\AA$^2$)\\
\hline
Ca  & 0.7653(5) & 0.2500 & 0.8683(5) & 0.719\\
Mn  & \textcolor{red}{0.2542(4)} & 0.2500 & 0.0000 & 0.714\\
Bi1 & \textcolor{red}{0.2494(1244}) & 0.2500 & 0.2500 & 0.620\\
Bi2 & 0.7328(3) & 0.2500 & 0.0806(2) & 0.462\\
\end{tabular}
\end{ruledtabular}
\begin{ruledtabular}
\begin{tabular}{ccccc}
& \multicolumn{3}{c}{ NSF + SF } & $\chi^2 = 3.85$\\
\hline
Atom & $x$ & $y$ & $z$ &$B_\mathrm{iso}$ (\AA$^2$)\\
\hline
Ca  & 0.7665(7) & 0.2500 & 0.8691(5) & 0.719\\
Mn  & \textcolor{red}{0.2537(4)} & 0.2500 & 0.0000 & 0.714\\
Bi1 & \textcolor{red}{0.2484(618}) & 0.2500 & 0.2500 & 0.620\\
Bi2 & 0.7316(3) & 0.2500 & 0.0805(2) & 0.462\\
\hline
\multicolumn{5}{c}{$\mu_{Mn}(x) = N/A$, $\mu_{Mn}(z) = 4.23(7)\mu_B$} \\
\end{tabular}
\end{ruledtabular}
\end{table}

For the $Z^{-}_5$ IR, the atomic positions obtained from the NSF-only and NSF+SF refinements are very similar, except for Mn, which shows markedly different $a$-axis shifts. This discrepancy suggests a notable trade-off between structural and magnetic contributions, indicating that the refinement is unreliable given the limited number of reflections measured. As discussed in the main text, spin canting cannot be resolved.

For the $Z^{+}_5$ IR, the refinements reveal similar but generally larger $a$-axis shifts (relative to $Z^{-}_5$) for most atoms, and spin canting is also not resolved. The large positional uncertainty of the sheared Bi1 layer suggests either a poorly constrained refinement or a structural inconsistency. The quality of the $Z^{+}_5$ refinement, as indicated by $\chi^2$, is slightly inferior to that of $Z^{-}_5$. In both cases broad, nearly equal-depth $\chi^2$ minima indicate that the refinements are poorly constrained.

While one reason for the poorly constrained refinements is the limited number of measured Bragg peaks, another is that a combined refinement of the lattice and magnetic structures using the total NSF+SF intensities is effectively equivalent to that of unpolarized data. This approach discards important information provided by polarized neutron diffraction, namely the separation of magnetic and nuclear contributions to Bragg intensities, which imposes additional constraints on both lattice and magnetic structure models.

\begin{figure}[h!]
\centering
\includegraphics[width=0.9\columnwidth]{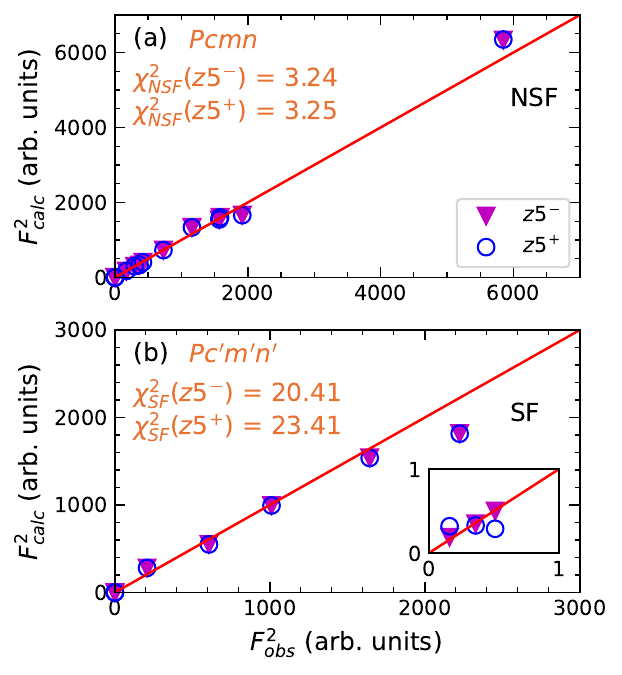}
\caption{Simultaneous refinements of the lattice and magnetic structures of \cmb\ using NSF and SF Bragg intensities measured at 5~K. The lattice refinement (a) was performed with crystal symmetry $Pcmn$, and the magnetic refinement (b) with magnetic symmetry $Pc'm'n'$. Calculated and observed intensities are compared as squared structure factors (arbitrary units). Magenta triangles denote $Z^{-}_5$ and blue open circles $Z^{+}_5$. The inset zooms in on weak superlattice peaks.
\label{TAS_simultaneous_fixBiso}}
\end{figure}

\begin{table}[h!]
\caption{\textbf{TAS simultaneous NSF and SF refinements.}
Crystal and magnetic structure parameters of \cmb\ at 5~K obtained from the simultaneous refinement of integrated NSF and SF Bragg intensities measured with polarized neutron diffraction. $B_\mathrm{iso}$ were fixed to values obtained from HB-3A refinements. Nuclear and magnetic parameters were listed together because the simultaneous refinement used one set of structural parameters. Lattice parameters: $a = 4.483$~\AA, $b = 4.481$~\AA, $c = 21.968$~\AA. \label{TAS_simultaneous_refinement}}

\begin{ruledtabular}
\begin{tabular}{ccccc}
$Z^{-}_5$ & \multicolumn{2}{c}{$\chi^{2}_{NSF} = 3.24$} & \multicolumn{2}{c}{$\chi^{2}_{SF} = 20.41$}\\
\hline
Atom & $x$ & $y$ & $z$ &$B_\mathrm{iso}$\\
\hline
Ca  & 0.7712(3) & 0.2500 & 0.6183(4) & 0.593\\
Mn  & 0.2535(2) & 0.2500 & 0.7500 & 0.713\\
Bi1 & \textcolor{red}{0.2464(1)} & 0.2500 & 0.5000 & 0.637\\
Bi2 & 0.7593(2) & 0.2500 & 0.8306(1) & 0.609\\
\hline
\multicolumn{5}{c}{$\mu_{\mathrm{Mn}}(x) = N/A$, $\mu_{\mathrm{Mn}}(z) = 3.82(3) \mu_B$}\\
\end{tabular}
\end{ruledtabular}
\begin{ruledtabular}
\begin{tabular}{ccccc}
$Z^{+}_5$ & \multicolumn{2}{c}{$\chi^{2}_{NSF} = 3.25$} & \multicolumn{2}{c}{$\chi^{2}_{SF} = 23.41$}\\
\hline
Atom & $x$ & $y$ & $z$ &$B_\mathrm{iso}$\\
\hline
Ca  & 0.7656(4) & 0.2500 & 0.8685(4) & 0.719\\
Mn  & \textcolor{red}{0.2543(3)} & 0.2500 & 0.0000 & 0.714\\
Bi1 & \textcolor{red}{0.2487(474}) & 0.2500 & 0.2500 & 0.620\\
Bi2 & 0.7326(2) & 0.2500 & 0.0806(1) & 0.462\\
\hline
\multicolumn{5}{c}{$\mu_{\mathrm{Mn}}(x) = 0.062(3) \mu_B$, $\mu_{\mathrm{Mn}}(z) = 3.81(3) \mu_B$}\\
\end{tabular}
\end{ruledtabular}
\end{table}

To better constrain our results, we carried out simultaneous refinement by comparing the NSF and SF intensities directly to the corresponding lattice and magnetic structure models for each IR ($Z^{-}_5$ and $Z^{+}_5$), using a single set of structural parameters (see Section~\ref{Chi2-opt} for optimization details). The ADPs $B_\mathrm{iso}$ were again fixed to the values obtained from unpolarized neutron refinements at the HB-3A DEMAND diffractometer (App.~\ref{App:HB3A_refinement}). Figure~\ref{TAS_simultaneous_fixBiso} summarizes these refinements, and the refined parameters are listed in Table~\ref{TAS_simultaneous_refinement}.

The simultaneous refinements suggest a preference for the $Z^{-}_5$ structure over $Z^{+}_5$, but still yield unreasonably large uncertainties in atomic positions, particularly for $Z^{+}_5$, as well as larger apparent $a$-axis shifts, which likely reflect a trade-off between NSF and SF contributions discussed earlier. Measurable spin canting in the $Z^{+}_5$ model is required to reproduce the superlattice magnetic intensity observed in the SF data.

These results show that even the full information available from the polarized measurements is still insufficient to obtain a reliable structural refinement, given the limited number of measured Bragg peaks. This led us to perform an additional unpolarized neutron refinement on the HB-3A DEMAND diffractometer.

\section{Structural refinement from four-circle unpolarized neutron diffraction}
\label{App:HB3A_refinement}

Single-crystal neutron diffraction measurements were performed in the four-circle mode of the DEMAND (HB-3A) diffractometer at the High Flux Isotope Reactor, Oak Ridge National Laboratory. A crystal of mass $m \approx 160$~mg was mounted on an aluminum holder in a closed-cycle refrigerator (CCR) with a base temperature of 5~K and aligned in the $(H, K, 0)$ horizontal scattering plane. Neutrons with a wavelength of 1.003~\AA, selected by a Si (331) monochromator, were used. The nuclear and magnetic structures were refined with the FullProf Suite using integrated intensities of the measured Bragg reflections.

\begin{figure}[h!]
\centering
\includegraphics[width=0.9\columnwidth]{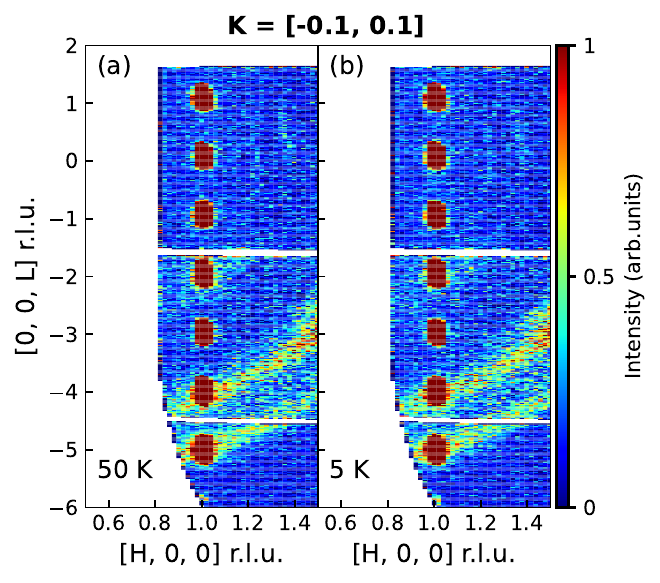}
\caption{Neutron diffraction intensity in the $(H,0,L)$ plane measured on the HB-3A DEMAND diffractometer at 50~K (a) and 5~K (b). The signal was integrated for $K \in [-0.1,0.1]$ reciprocal lattice units (r.l.u.). Lattice indexing corresponds to the tetragonal $P4/nmm$ structure at 50~K. Bright spots at $H = 1$ are lattice Bragg peaks, while weaker scattering bands at the bottom originate from diffraction by polycrystalline aluminum in the sample environment. \label{HB3A_contour}}
\end{figure}

Figure~\ref{HB3A_contour} shows contour plots of neutron diffraction intensity slices through the $(H,0,L)$ plane, integrated over $K \in [-0.1,0.1]$, measured on the HB-3A diffractometer at 50~K and 5~K. The $L = 2n + \tfrac{1}{2}$ superlattice peaks (indexed in tetragonal $P4/nmm$) are not visible because they are too weak to be detected in this measurement. Likely for this reason, refinement of the low-temperature $Pcmn$ structure, which is only slightly distorted relative to $P4/nmm$, does not converge even with the constrained ADPs. As summarized in Table~\ref{Table:HB3A_50K_5K}, both the 50~K and 5~K datasets can be equally well refined using the high-temperature tetragonal $P4'/n'm'm$ structure without $c$-axis doubling, which does not account for the superlattice reflections.

\begin{table}[]
\caption{\textbf{HB-3A refinement at 50~K and 5~K.} Crystal and magnetic structure parameters of \cmb\ at 50~K and 5~K obtained from Rietveld refinement of integrated Bragg intensities measured with neutron diffraction on HB-3A DEMAND assuming high-temperature tetragonal $P4'/n'm'm$ structure, which does not account for the 5~K superlattice reflections. Lattice parameters: $a = b = 4.47$~\AA, $c = 10.89$~\AA. \label{Table:HB3A_50K_5K}}
\begin{ruledtabular}
\begin{tabular}{ccccc}
T = 50~K & \multicolumn{2}{c}{$P4'/n'm'm$} & \multicolumn{2}{c}{$\chi^{2} = 5.66$}\\
\hline
Atom & $x$ & $y$ & $z$ &$B_\mathrm{iso}$ (\AA$^2$)\\
\hline
Ca  & 0.2500 & 0.2500 & 0.7846(67) & 0.609(97)\\
Mn  & 0.7500 & 0.2500 & 0.0000 & 0.224(119)\\
Bi1 & 0.7500 & 0.2500 & 0.5000 & 0.681(62)\\
Bi2 & 0.7500 & 0.7500 & 0.8028(37) & 0.713(60)\\
\hline
\multicolumn{5}{c}{$\mu_{\mathrm{Mn}}(x) = N/A$, $\mu_{\mathrm{Mn}}(z) = 3.68(5) \mu_B$}\\
\end{tabular}
\end{ruledtabular}
\begin{ruledtabular}
\begin{tabular}{ccccc}
T = 5~K & \multicolumn{2}{c}{$P4'/n'm'm$} & \multicolumn{2}{c}{$\chi^{2} = 4.27$}\\
\hline
Atom & $x$ & $y$ & $z$ &$B_\mathrm{iso}$ (\AA$^2$)\\
\hline
Ca  & 0.2500 & 0.2500 & 0.7252(26) & 0.868(123)\\
Mn  & 0.7500 & 0.2500 & 0.0000 & 0.724(146)\\
Bi1 & 0.7500 & 0.2500 & 0.5000 & 0.644(83)\\
Bi2 & 0.7500 & 0.7500 & 0.8402(12) & 0.669(90)\\
\hline
\multicolumn{5}{c}{$\mu_{\mathrm{Mn}}(x) = N/A$, $\mu_{\mathrm{Mn}}(z) = 3.72(7) \mu_B$}\\
\end{tabular}
\end{ruledtabular}
\end{table}

\begin{table}[b!]
\caption{\textbf{$T =$ 5~K refinement.} Crystal and magnetic structure parameters of \cmb\ at 5~K obtained from Rietveld refinement of integrated Bragg intensities measured with neutron diffraction on DEMAND, including three re-scaled superlattice peaks from TAS measurements, without spin canting. Lattice parameters: $a = 4.456$~\AA, $b = 4.459$~\AA, $c = 21.158$~\AA. \label{Table:HB3A_HB1_5K_z5m_z5p} }
\begin{ruledtabular}
\begin{tabular}{ccccc}
T = 5~K & \multicolumn{2}{c}{$Pc^\prime m^\prime n^\prime$} $Z^{-}_5$& \multicolumn{2}{c}{$\chi^{2} = 4.13$}\\
\hline
Atom & $x$ & $y$ & $z$ &$B_\mathrm{iso}$ (\AA$^2$)\\
\hline
Ca  & 0.7689(20) & 0.2500 & 0.6124(10) & 0.593(76)\\
Mn  & 0.2526(47) & 0.2500 & 0.7500 & 0.713(63)\\
Bi1 & \textcolor{red}{0.2466(11)} & 0.2500 & 0.5000 & 0.637(28)\\
Bi2 & 0.7580(11) & 0.2500 & 0.8299(5) & 0.609(36)\\
\hline
\multicolumn{5}{c}{$\mu_{\mathrm{Mn}}(x) = N/A$, $\mu_{\mathrm{Mn}}(z) = 3.73(4) \mu_B$}\\
\end{tabular}
\end{ruledtabular}
\begin{ruledtabular}
\begin{tabular}{ccccc}
T = 5~K & \multicolumn{2}{c}{$Pc^\prime m^\prime n^\prime$} $Z^{+}_5$& \multicolumn{2}{c}{$\chi^{2} = 4.16$}\\
\hline
Atom & $x$ & $y$ & $z$ &$B_\mathrm{iso}$ (\AA$^2$)\\
\hline
Ca  & 0.7639(20) & 0.2500 & 0.8624(10) & 0.719(76)\\
Mn  & \textcolor{red}{0.2527(17)} & 0.2500 & 0.0000 & 0.714(63)\\
Bi1 & 0.2445(32) & 0.2500 & 0.2500 & 0.620(28)\\
Bi2 & 0.7342(9) & 0.2500 & 0.0799(5) & 0.462(36)\\
\hline
\multicolumn{5}{c}{$\mu_{\mathrm{Mn}}(x) = N/A$, $\mu_{\mathrm{Mn}}(z) = 3.74(4) \mu_B$}\\
\end{tabular}
\end{ruledtabular}
\end{table}

\begin{figure}[]
\centering
\includegraphics[width=0.9\columnwidth]{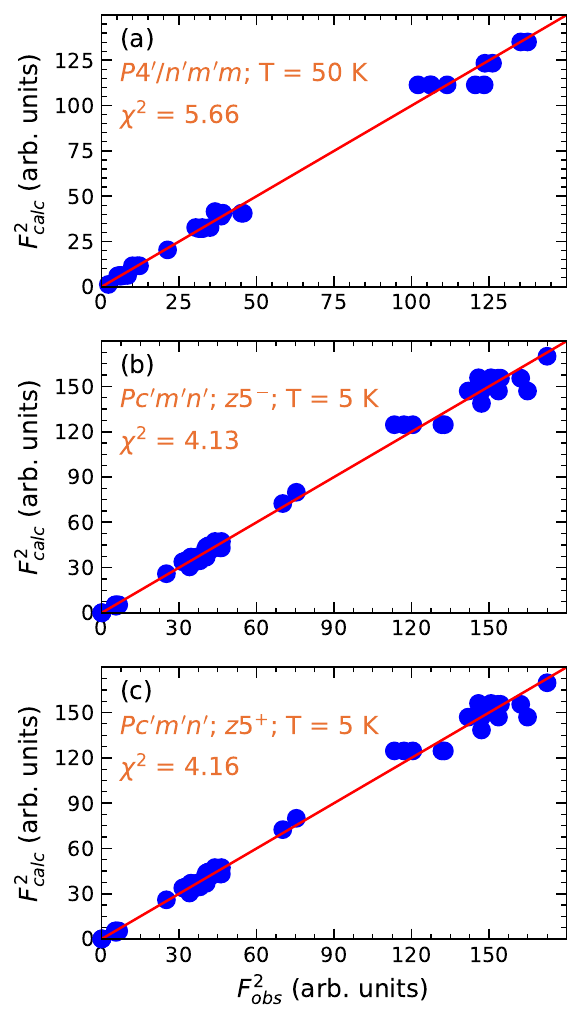}
\caption{Summary of the structural and magnetic Rietveld refinements of \cmb: (a) using HB-3A data only at 50~K with $P4/nmm$ symmetry, and (b) using the HB-3A data supplemented with the three re-scaled superlattice peaks measured on HB-1 at  at 5~K in (b) $Z^{-}_5$ and (c) $Z^{+}_5$ IRs with magnetic symmetry $Pc^\prime m^\prime n^\prime$ (standard setting $Pn^\prime m^\prime a^\prime$). The plots compare calculated and observed intensities, expressed as squared structure factors in arbitrary units.}
\label{HB3A_refinement_5K_50K}
\vspace{-0.25in}
\end{figure}

To address this limitation, we repeated the HB-3A refinement of the lattice and magnetic structures at $T = 5$~K, supplementing the dataset with three superlattice peaks, $(1,0,0.5)$, $(1,0,1.5)$, and $(1,0,2.5)$, observed on the HB-1 triple-axis spectrometer. Their intensities were re-scaled using the average ratio of the four $(H,0,L)$ main lattice Bragg peaks ($H = 1$, $L = 0,2,4,6$ in $Pcmn$ notation) measured on both instruments, $\tfrac{I(\mathrm{HB\text{-}1})}{I(\mathrm{HB\text{-}3A})} = 42.585$. The refinements of the HB-3A dataset supplemented with the three re-scaled HB-1 superlattice peaks at 5~K are summarized in Table~\ref{Table:HB3A_HB1_5K_z5m_z5p} and shown in Fig.~\ref{HB3A_refinement_5K_50K}, together with the $P4'/n'm'm$ refinement of the 50~K data.

The refinement of the low-temperature $Pcmn$ structure at 5~K using the HB-3A dataset supplemented with the three re-scaled HB-1 superlattice peaks converges, indicating a slight preference for the $Z^{-}_5$ model. However, broad or multiple $\chi^2$ minima obtained when varying the zigzag displacements of Bi1 ($Z^{-}_5$) or Mn ($Z^{+}_5$) along the $a$-axis show that these parameters remain poorly constrained, similar to the TAS-only refinements discussed in Sec.~\ref{App:TAS_refinement}. We therefore performed a combined refinement of the HB-3A and polarized TAS datasets, the results of which are presented in the main manuscript. The corresponding optimization procedures are described below in Sec.~\ref{Chi2-opt}.

Finally, to assess the stoichiometry of our crystal, we performed sequential refinements of the site occupancies (occ) for all four Wyckoff sites using the HB-3A data. The results [occ(Ca) = 0.1232(63), occ(Mn) = 0.1236(68), occ(Bi1) = 0.1252(33), and occ(Bi2) = 0.1266(36)] indicate that the sample composition is stoichiometric within $\approx 1.4\%$, albeit with a statistical uncertainty of $\approx 5\%$.

\section{{Calculated and Experimental Intensities from Polarized Neutron Diffraction Refinement}}
\label{App:PND_refinement}

{Table~\ref{Iobs_vs_Ical} lists the observed non-spin-flip (NSF) and spin-flip (SF) intensities from polarized neutron diffraction, together with the corresponding calculated values from the refinements discussed in Sec.~\ref{PND} and Apps.~\ref{App:TAS_refinement} and \ref{App:HB3A_refinement}. No measurable magnetic superlattice SF intensity was detected for even-$L$ peaks. The small SF intensity at (2,0,2) originates from an Al-powder contamination that is not fully corrected using the single-crystal Bragg flipping ratio and was therefore excluded from the refinements. $I_{\mathrm{obs}}$ are the original TAS NSF and SF intensities. The table also lists TAS intensities rescaled to the HB-3A measurement by a factor of 42.585, as noted in the manuscript. These rescaled intensities are used for, and should be compared with, the results of the combined refinement.}

{For $(2,0,L=\mathrm{even})$ reflections, the SF intensity is symmetry-allowed in the presence of spin canting for both $Z_5^{-}$ and $Z_5^{+}$ [${\sim}M_x\sin(4\pi\delta)$]. In our refinements, however, $Z_5^{-}$ has no canting ($M_x=0$), while $Z_5^{+}$ yields only a very small $M_x$, producing negligible SF intensity ($<10^{-5}$, below the numerical accuracy of \textsc{FullProf}).}

\section{Optimization procedure for the combined refinement}
\label{Chi2-opt}

\begin{table*}[t!]
\centering
\caption{{Observed and calculated intensities, $I_{\mathrm{obs}}$ and $I_{\mathrm{cal}}$, from the polarized-neutron-diffraction refinements for the non-spin-flip (NSF; top) and spin-flip (SF; bottom) channels. NSF $I_{\mathrm{cal}}$ are shown for the NSF-only refinement, NSF-only with fixed $B_{\mathrm{iso}}$ ADPs, simultaneous (NSF,SF) refinement, and the combined refinement (polarized TAS + unpolarized HB3A). SF $I_{\mathrm{cal}}$ are shown for the simultaneous (NSF,SF) and combined refinements. Entries where the $Z_5^{+}$ model shows pronounced discrepancies are highlighted in red.
Peaks are indexed in the low-temperature orthorhombic $Pcmn$ cell used for the refinements; in the high-temperature $P4/nmm$ indexing used in the main text, the present $L$ index is divided by 2. }}
\label{Iobs_vs_Ical}
\begin{ruledtabular}
\begin{tabular}{c|r|r r|r r|r r|r|r r}
\multicolumn{11}{c}{NSF intensity} \\
\hline
Peak & NSF & \multicolumn{2}{c|}{NSF-only} & \multicolumn{2}{c|}{NSF-only fix Biso} & \multicolumn{2}{c|}{(NSF, SF) simultaneous} & {Rescaled} & \multicolumn{2}{c}{Combined} \\
(H,K,L)
& {$I_{\mathrm{obs}}$}
& {$Z_5^{-}$} & {$Z_5^{+}$}
& {$Z_5^{-}$} & {$Z_5^{+}$}
& {$Z_5^{-}$} & {$Z_5^{+}$}
& {$I_{\mathrm{obs}}$}
& {$Z_5^{-}$} & {$Z_5^{+}$} \\
\hline
(1,0,1) & 2.21   & 2.21   & 2.21   & 2.21   & 2.21   & 2.21   & 2.21   & 0.052 & 0.052 & 0.054 \\
(1,0,2) & 172.53 & 175.83 & 175.99 & 178.64 & 176.89 & 173.07 & 173.37 & 4.051 & 4.094 & 4.125 \\
(1,0,3) & 1.26   & 1.26   & 1.26   & 1.26   & 1.26   & 1.26   & 1.26   & 0.030 & 0.030 & 0.030 \\
(1,0,4) & 1917.95& 1670.54& 1669.20& 1654.65& 1645.39& 1664.98& 1660.42& 45.038& 41.308& 41.280 \\
(1,0,5) & 5.02   & 5.02   & 5.02   & 5.02   & 5.03   & 5.02   & 5.03   & 0.118 & 0.117 & 0.115 \\
(1,0,6) & 731.46 & 738.82 & 739.44 & 742.98 & 740.11 & 730.47 & 730.68 & 17.176& 18.531& 18.530 \\
(1,0,8) & 1584.09& 1603.71& 1603.52& 1620.38& 1607.54& 1605.97& 1603.96& 37.198& 39.900& 39.946 \\
(0,0,2) & 1160.49& 1297.18& 1294.07& 1270.60& 1266.56& 1339.26& 1335.55& 27.251& 32.270& 32.188 \\
(0,0,4) & 371.76 & 350.03 & 348.48 & 338.55 & 335.30 & 331.58 & 333.34 & 8.730 & 7.676 & 7.712 \\
(0,0,6) & 5847.61& 5642.12& 5650.34& 5803.37& 5829.63& 6331.57& 6346.28& 137.316&154.044&154.350 \\
(0,0,8) & 422.66 & 422.02 & 421.73 & 417.88 & 406.87 & 408.61 & 405.94 & 9.925 & 9.502 & 9.430 \\
(2,0,2) & 1572.76& 1571.37& 1573.17& 1529.95& 1545.62& 1542.88& 1547.19& 36.932& 38.577& 38.584 \\
(2,0,4) & 297.94 & 297.96 & 298.27 & 300.74 & 307.29 & 302.27 & 302.65 & 6.996 & 7.101 & 7.119 \\
\bottomrule
\end{tabular}
\begin{tabular}{c|r|r r|r|r r}
\multicolumn{7}{c}{SF intensity} \\
\hline
Peak & SF & \multicolumn{2}{c|}{(NSF, SF) simultaneous} & {Rescaled} & \multicolumn{2}{c}{Combined} \\
(H,K,L)& {$I_{\mathrm{obs}}$}
& {$Z_5^{-}$} & {$Z_5^{+}$}
& {$I_{\mathrm{obs}}$}
& {$Z_5^{-}$} & {$Z_5^{+}$} \\
\hline
(1,0,0) & 2224.51 & 1817.77 & 1815.05 & 52.237 & 46.893 & 46.852 \\
(1,0,1) &    0.51 &    0.50 &    \textcolor{red}{0.29} &  0.012 &  0.012 &  \textcolor{red}{0.007} \\
(1,0,2) & 1643.94 & 1539.84 & 1538.05 & 38.604 & 39.442 & 39.432 \\
(1,0,3) &    0.36 &    0.35 &    0.33 &  0.008 &  0.008 &  0.008 \\
(1,0,4) & 1010.77 &  995.18 &  994.64 & 23.735 & 25.032 & 25.051 \\
(1,0,5) &    0.16 &    0.19 &    \textcolor{red}{0.32} &  0.004 &  0.004 &  \textcolor{red}{0.008} \\
(1,0,6) &  607.02 &  549.78 &  549.72 & 14.254 & 13.510 & 13.528 \\
(1,0,8) &  207.58 &  280.86 &  280.89 &  4.874 &  6.734 &  6.743 \\

(0,0,2) &  0  &  0  &  0  &  0  &  0  &  0 \\
(0,0,4) &  0  &  0  &  0  &  0  &  0  &  0 \\
(0,0,6) &  0  &  0  &  0  &  0  &  0  &  0 \\
(0,0,8) &  0  &  0  &  0  &  0  &  0  &  0 \\
(2,0,2) & 10  &  0  &  \textless $10^{-5}$  &  0.235  &  0  &  \textless $10^{-6}$ \\
(2,0,4) &  0  &  0  &  \textless $10^{-5}$  &  0      &  0  &  \textless $10^{-6}$ \\
\end{tabular}
\end{ruledtabular}
\end{table*}

For NSF-only, (NSF+SF), and unpolarized HB3A refinements, we optimize the usual normalized $\chi^2$ function:
\begin{equation} \label{eq1}
\begin{split}
\chi^2 = \frac{1}{N}\sum_{n=1}^{N}\frac{(I^{obs}_n - I^{calc}_n)^2}{\sigma_n^2},
\end{split}
\end{equation}
where $I^{obs}_n$ and $\sigma_n$ are the observed intensity and its uncertainty at point $n$, and $I^{calc}_n$ is the corresponding intensity the calculated from structural model.

For the simultaneous NSF and SF refinements, we optimize the combined $\chi^2$ function:
\begin{equation} \label{eq1}
\begin{split}
\chi^2 = \frac{\left[ \sum_{n=1}^{N_{NSF}} \left(\frac{I^{obs}_n - I^{calc}_n}{\sigma_n}\right)^2 + \sum_{n=1}^{N_{SF}} \left(\frac{I^{obs}_n - I^{calc}_n}{\sigma_n}\right)^2
\right]}{N_{NSF} + N_{SF}}
\end{split}
\end{equation}

Similarly, for the combined unpolarized and polarized data refinement, we optimize the combined $\chi^2$ function (unpolarized intensities include those measured at HB-3A and the re-scaled (NSF+SF) intensities from the polarized TAS measurement, which add the superlattice peaks otherwise unobservable at HB-3A):
\begin{equation} \label{eq1}
\begin{split}
\chi^2 = \frac{\sum_{type}^{unpolar, NSF,SF}
\left[ \sum_{n}^{N_{type}} \left( \frac{I^{obs}_n - I^{calc}_n}{\sigma_n} \right)^2 \right] }
{N_{unpolar} + N_{NSF} + N_{SF}}
\end{split}
\end{equation}

\section{Details of DFT Analysis}
\label{DFT_appendix}

\begin{figure*}[t!]
\centering
\begin{overpic}[width=0.3\textwidth]{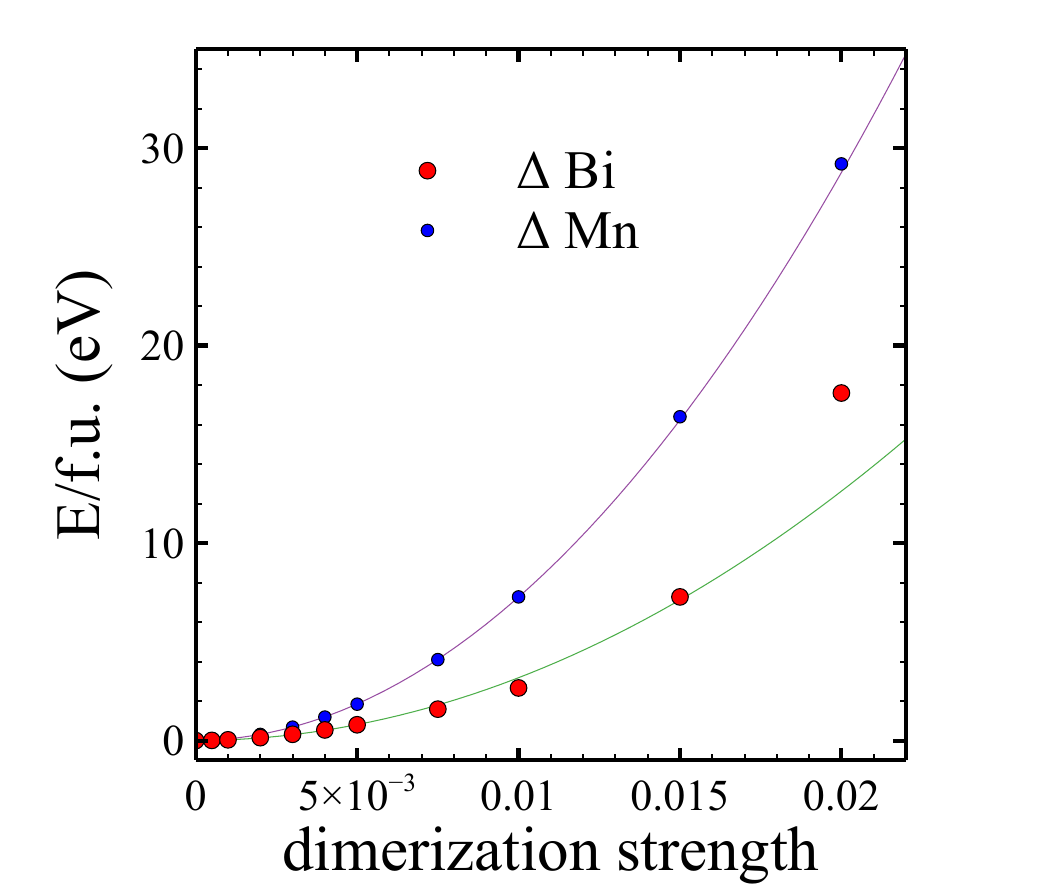}
    \put(5,80){(a)}
\end{overpic}
\hfill
\begin{overpic}[width=0.34\textwidth]{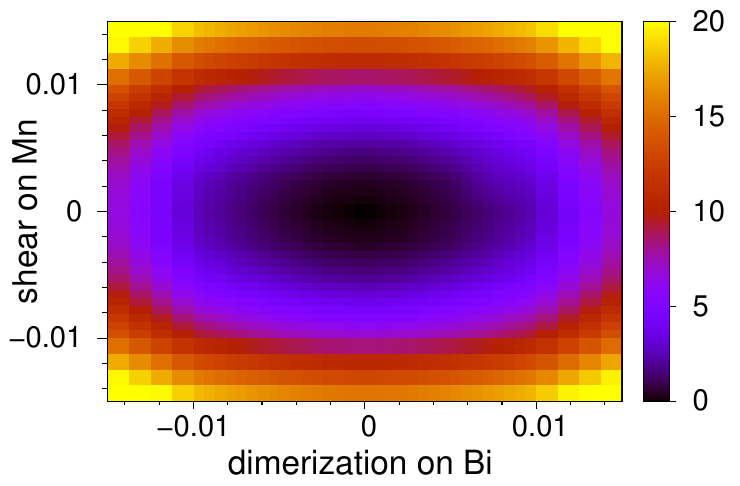}
    \put(5,70){(b)}
\end{overpic}
\hfill
\begin{overpic}[width=0.34\textwidth]{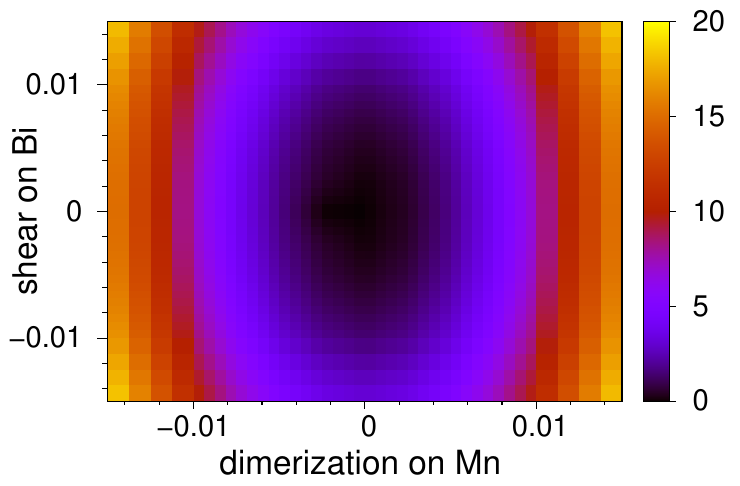}
    \put(5,70){(c)}
\end{overpic}
\caption{(a) Energy cost per formula unit for bond dimerization on Bi and Mn square layers using fully relaxed structures. Dots represent DFT data points, and lines are quadratic fits.
(b) Two-dimensional energy landscape as a function of dimerization strength on the Bi layer and shear distortion on the Mn layer.
(c) Similar to panel (b), but with shear on the Bi layer and dimerization on the Mn layer. Dimerization and shear amplitudes are given in \AA.}
\label{Fig:DFT_Fig1S_energy}
\end{figure*}

\begin{figure*}[t!]
\centering
\begin{overpic}[width=0.54\textwidth]{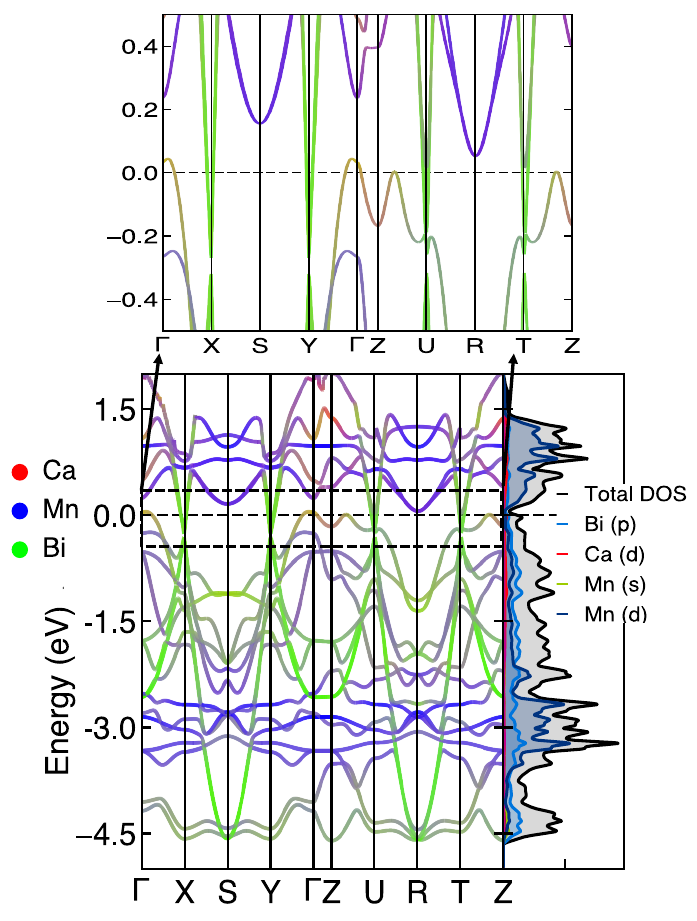}
    \put(5,95){\textbf{(a)}}
\end{overpic}
\hfill
\begin{overpic}[width=0.45\textwidth]{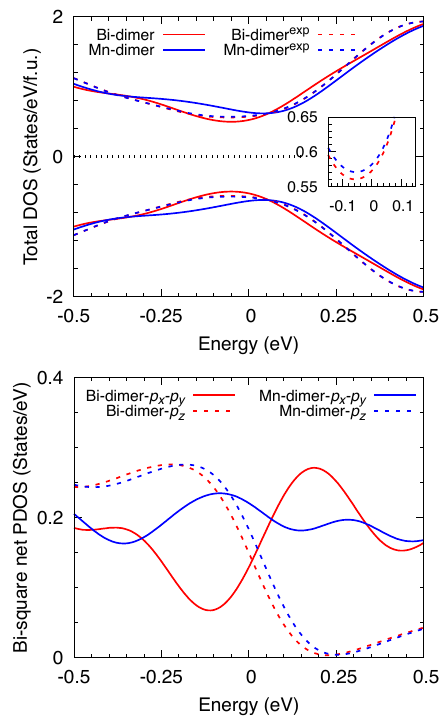}
    \put(0,95){\textbf{(b)}}
\end{overpic}
\caption{\label{Fig:DFT_Fig2S_bands}(a) Electronic band structure and density of states (DOS) for the undistorted tetragonal phase, based on the DFT-relaxed structure. Colors indicate contributions from different atoms. The inset shows a zoomed view of the Dirac feature within a $\pm$0.5~eV energy window. (b) Comparison of the total spin-resolved DOS (top panel) and partial DOS from Bi square-net 5$p$ states (bottom panel) for various dimerized phases. The DOS from experimentally refined Bi- and Mn-dimerized phases (see inset) are nearly indistinguishable. To highlight spectral differences, we artificially increased the dimerization strength by 0.015~\AA. The lower panel reveals that the primary change in DOS arises from redistribution of spectral weight among Bi $p_x$ and $p_y$ orbitals in the square net. }
\end{figure*}

\subsection{Computational Details}
\label{App:DFT_details}
Density functional theory (DFT) calculations were performed using the Vienna \textit{Ab initio} Simulation Package (VASP) \cite{Kresse_VASP1_PRB1993,Bloechl_PAW_PRB1994}. The Perdew–Burke–Ernzerhof (PBE) exchange-correlation functional~\cite{Perdew_PBE_PRL1996}, within the generalized gradient approximation (GGA), was used for all calculations. A plane-wave energy cutoff of 500~eV and a Monkhorst–Pack $k$-point mesh of $15 \times 15 \times 7$ were used to sample the Brillouin zone of the primitive cell. Lattice relaxation calculations were carried out with electronic and ionic energy convergence criteria of $10^{-7}$~eV and $10^{-4}$~eV, respectively. Spin–orbit coupling was not included in these calculations.

\subsection{Total Energy of Competing Lattice Distortions}
\label{DFT_Total_Energy}
Since the energy scales involved are very small and our full lattice relaxation calculations did not yield a stable dimerized phase, we performed additional total energy calculations exploring the space of shear and dimerization amplitudes. Figure~\ref{Fig:DFT_Fig1S_energy}(a) shows the energy cost for varying dimerization amplitudes in the Mn and Bi layers without introducing shear distortion. Interestingly, we find that dimerization in the Bi square layer is energetically more favorable than in the Mn layer, although both configurations correspond to higher-energy states. This observation is consistent with the result in Table~\ref{Table:DFT}, which shows that the $Z_5^-$ structure is slightly lower in energy than the $Z_5^+$ structure. In Figs.~\ref{Fig:DFT_Fig1S_energy}(b) and \ref{Fig:DFT_Fig1S_energy}(c), we present the total energy landscape as a function of both dimerization and shear distortions for the $Z_5^-$ and $Z_5^+$ configurations, respectively. In all cases, the minimum energy corresponds to the undistorted tetragonal structure in our calculations.

The orbital-resolved electronic band structure and the corresponding DOS obtained from our DFT calculations are presented in Fig.~\ref{Fig:DFT_Fig2S_bands}(a). To understand how the dimerized phases affect the electronic structure, we computed the DOS for the $Z_5^-$ and $Z_5^+$ configurations in a narrow energy window around the Fermi level, as shown in Fig.~\ref{Fig:DFT_Fig2S_bands}(b). As discussed in the main text, the DOS profiles for the experimentally refined dimerization amplitudes are nearly identical, which is expected given that the dimerization amplitude is very small ($\sim 10^{-3}$\AA). Upon closer inspection, however, the $Z_5^-$ phase exhibits a slightly reduced DOS at the Fermi level compared to $Z_5^+$. Since this difference lies well within the uncertainty of DFT, we performed additional calculations using a fully relaxed structure with an artificially enhanced dimerization strength of 0.015\AA\ to examine whether the Bi-dimerized phase indeed yields a lower DOS and exhibits signatures of a charge density wave (CDW) gap. With this increased dimerization, the Bi-dimerized $Z_5^-$ phase shows a clear reduction in the DOS near the Fermi level due to a redistribution of electronic population within the Bi $p_x$–$p_y$ orbitals [see Fig.~\ref{Fig:DFT_Fig2S_bands}(b), lower panel], consistent with the fact that the states near the Fermi level in the undistorted tetragonal phase are primarily derived from Bi 5\textit{p} orbitals in the square-net layer.

\end{document}